\documentclass[fleqn,usenatbib]{mnras}

\usepackage{newtxtext,newtxmath}

\usepackage[T1]{fontenc}
\usepackage{CJKutf8}
\usepackage[normalem]{ulem}

\DeclareRobustCommand{\VAN}[3]{#2}
\let\VANthebibliography\thebibliography
\def\thebibliography{\DeclareRobustCommand{\VAN}[3]{##3}\VANthebibliography}

\usepackage{graphicx}	
\usepackage{amsmath}	
\newcommand{\de}{\operatorname{d}}
\newcommand{\HOfromTOandSNeBF}{\left({74.7\pm 9.6}\right) \ \mathrm{km/s/Mpc}}
\newcommand{\HOfromTOandBAOBF}{\left({72.9^{+10.0}_{-8.6}}\right) \ \mathrm{km/s/Mpc}}

\title[eBOSS: Measurement of the equality scale]{Measurement of the matter-radiation equality scale using the extended Baryon Oscillation Spectroscopic Survey Quasar Sample}

\author[B. Bahr-Kalus, D. Parkinson and E.-M. Mueller]{
Benedict Bahr-Kalus$^{1}$\thanks{E-mail: benedictbahrkalus@kasi.re.kr}, David Parkinson$^{1,2}$\thanks{E-mail: davidparkinson@kasi.re.kr} and Eva-Maria Mueller$^{3}$\thanks{E-mail: eva-maria.mueller@port.ac.uk}
\\
$^{1}$Korea Astronomy and Space Science Institute, Yuseong-gu, Daedeok-daero 776, Daejeon 34055, Republic of Korea\\
$^{2}$University of Science and Technology, Daejeon 34113, Republic of Korea\\
$^{3}$Institute of Cosmology and Gravitation, Dennis Sciama Building, University of Portsmouth, Portsmouth PO1 3FX, UK}

\date{Accepted XXX. Received YYY; in original form ZZZ}

\pubyear{2023}

\begin{document}
\label{firstpage}
\pagerange{\pageref{firstpage}--\pageref{lastpage}}
\maketitle

\begin{abstract}
The position of the peak of the matter power spectrum, the so-called turnover scale, is set by the horizon size at the epoch of matter-radiation equality. It can easily be predicted in terms of the physics of the Universe in the relativistic era, and so can be used as a standard ruler, independent of other features present in the matter power spectrum, such as baryon acoustic oscillations (BAO). We use the distribution of quasars measured by the extended Baryon Oscillation Spectroscopic Survey (eBOSS) to determine the turnover scale in a model-independent fashion statistically. We avoid modelling the BAO by down-weighting affected scales in the covariance matrix using the mode deprojection technique. We measure the wavenumber of the peak to be $k_\mathrm{TO} = \left( 17.6^{+1.9}_{-1.8} \right) \times 10^{-3}h/\mathrm{Mpc}$, corresponding to a dilation scale of $ D_\mathrm{V}(z_\mathrm{eff} = 1.48) = \left({36.2^{+4.1}_{-4.4}}\right)r_\mathrm{H}$. This is not competitive with current BAO distance measures in terms of determining the expansion history but does provide a useful cross-check. We combine this measurement with low-redshift distance measurements from type-Ia supernova data from Pantheon and BAO data from eBOSS to make a sound-horizon free estimate of the Hubble-Lema\^itre parameter and find it to be  $H_0=\HOfromTOandSNeBF$ with Pantheon, and $H_0=\HOfromTOandBAOBF$ with eBOSS BAO.
We make predictions for the measurement of the turnover scale by the Dark Energy Spectroscopic Instrument (DESI) survey, the Maunakea Spectroscopic Explorer (MSE) and MegaMapper, which will make more precise and accurate distance determinations.
\end{abstract}


\section{Introduction}
\label{sec:intro}

The large-scale distribution of matter in the Universe can be decomposed into fluctuations. The amplitudes of these fluctuations as a function of their wavenumber are described by the matter power spectrum, $P(k)$. This matter power spectrum contains valuable information that can be used to determine the nature of the physics of the universe, including the baryon acoustic oscillation (BAO) standard rulers \citep{SDSS:2005xqv,2dFGRS:2005yhx,Beutler:2011hx,Kazin:2014qga,BOSS:2016wmc,duMasdesBourboux:2017mrl,Bautista:2020ahg,Gil-Marin:2020bct,Raichoor:2020vio,deMattia:2020fkb,Hou:2020rse,Neveux:2020voa}, and redshift-space distortions \citep[][and previously cited eBOSS references]{Kaiser:1987,2dFGRS:2004cmo,Blake:2011rj,Beutler:2012px,Okumura:2015lvp,Pezzotta:2016gbo,BOSS:2016wmc}. There is another standard ruler encoded in the matter power spectrum that is the position of the peak, also called the turnover scale.

The current best model of the initial conditions predicts adiabatic density fluctuations with a nearly scale-invariant spectrum, generated in the very early Universe. These density fluctuations experience a gravitational evolution dependent on the background dynamics of the expansion, their extent relative to the Hubble horizon size, and the presence of any pressure that can prevent gravitational collapse (as described by linear perturbation theory). Undergoing this processing over the whole history of the Universe, these fluctuations directly convey the large-scale matter power spectrum at late times, which we measure through the distribution of galaxies and other test particles such as quasars. 

However, there are large-scale fluctuations that have remained outside of the horizon since they were generated, and so exist on scales larger than the peak. These can give a window into the relativistic era, even when observed at late times {\citep[e.g.][]{Peter:2013avv}}. So far, because of the volume required to observe these very large-scale fluctuations with any statistical precision, this window is only starting to open and could be explored by the next generation of cosmological surveys.

The turnover scale is the point of transition between these unprocessed fluctuations, and those that re-entered the horizon before now and underwent significant evolution. It appears as a peak in the power spectrum and is fixed by the epoch at which the density of non-relativistic matter and relativistic particles (such as photons and neutrinos) were identical, the time of matter-radiation equality $t_{\rm eq}$.\footnote{Because of the cosmological uncertainty on the redshift-time relation, instead of using $t_{\rm eq}$, it is more common to use redshift as the coordinate here, and refer to $z_{\rm eq}$ instead, which we do for the remainder of the paper.}

Previous large-scale galaxy surveys have succeeded in detecting features in the power spectra on scales smaller than the turnover, such as the positions of the baryon acoustic oscillations, and the redshift-space distortions in the anisotropic distribution. However, direct turnover measurement
has only become possible recently, with surveys covering a cosmological volume in excess of $\sim$ 1 Gpc${^3}$. An attempt was made with the WiggleZ Dark Energy Survey data \citep{Poole:2012ex}, which provided a first tentative detection of the turnover (in a model-independent fashion). Still, it could not rule out power spectra that continued to rise with increasing wavenumber.

In this paper, we use the positions of quasars measured by eBOSS \citep{Ross:2020lqz} to reconstruct the large-scale power spectrum
in a model-independent fashion (similar to \citet{Poole:2012ex}), to detect and constrain the position of the turnover scale.  This kind of analysis is separate but complementary to full-shape measurements such as \citet{Vagnozzi:2020rcz}, \citet{Kobayashi:2021oud} and \citet*{Glanville:2022xes}. These full-shape analyses use the whole available range of the power spectrum to make inferences directly about the parameters of the cosmological model, and so are by definition model dependent \citep*[e.g.][]{Smith:2022}. 
On the plus side, this allows measurements where data is sparse. For instance, \citet{Philcox} make a statistical inference on the values of the equality scale and $H_0$ from BOSS DR12 power spectra \citep{BOSS:2016wmc}, but here unknown systematic effects mean that only a single data point is available below the expected equality scale \citep{Kalus:2018qsy}. On the downside,
the publicly available Effective Field Theory of Large-Scale Structure (EFTofLSS) codes \textsc{PyBird} and \textsc{CLASS-PT} provide discrepant constraints (at $\sim 0.9\sigma$) on $H_0$ from the same BOSS data \citep{Simon:2022lde}. Whereas this analysis seeks to extract only the position of the peak, and so encodes
the geometric information provided by the position while making as few assumptions about the cosmological model as possible. The \textsc{ShapeFit} extension \citep*{Brieden:2021edu} to classic BAO and RSD measurements adds information from the shape of the large-scale power spectrum in a model-independent way and has been applied to the complete set of BOSS and eBOSS data by \citet*{Brieden:2022lsd}. However, their approach is very different and complementary from the one we follow here: rather than measuring the horizon size at matter-radiation equality, the \textsc{ShapeFit} parameter determines how fast the horizon expanded at that epoch, producing different degeneracy directions in the $\Omega_\mathrm{m}$-$H_0$ plane. Combining the two approaches, \citet*{BriedenTale} therefore forecast improved constraints on the two parameters. In practice, this requires a careful analysis of the covariance between the two approaches, and, hence, we leave it for future work.

One merit of measuring the turnover scale lies in studying the so-called \textit{Hubble tension}. Recent local $H_0$-measurements using Cepheids and Supernovae on a distance ladder \citep[e.g. ][]{Riess:2021jrx} are in tension with the value inferred from \textit{Planck} \citep{Planck:2018vyg}. This tension persists even after carefully analysing the sample variance in $H_0$ \citep{Zhai:2022zif}. 
As the Hubble tension can be regarded as a sound-horizon tension \citep{Bernal:2016gxb,Aylor:2018drw,Knox:2019rjx}, one point of interest in studying the turnover scale lies in its use as a standard ruler, in fact, the longest in the Universe, that is also independent of the comoving sound horizon $r_\mathrm{drag}$ at the redshift at which the baryon-drag optical depth equalled unity. This is particularly interesting to test, for instance, early dark energy (EDE) models that solve the Hubble tension by introducing a component to the cosmic fluid that behaves like a cosmological constant at early times (redshifts
$z \ga 3000$) and then dilutes away before photon-decoupling \citep{Poulin:2018cxd}. Alternatively, dark energy and dark matter may interact non-gravitationally, which would also directly impact matter-radiation equality \citep{Nunes}. While we can measure $H_0$ with great precision and independently of the CMB by combining BAO with $r_\mathrm{drag}$ inferred from Big Bang Nucleosynthesis \citep[BBN; e.g. ][]{Schoeneberg}, BBN constraints rely on correctly modelling nuclear physics. As demonstrated in this work, combining the turnover point with low-redshift data provides a consistency check based only on simple physics.

The layout of this paper is as follows: in Sec.~\ref{sec:theory} we discuss the theory of the turnover position, and the model-independent parameterisation we will use to measure it. In section \ref{sec:data}, we review the eBOSS QSO data used to measure the turnover, and in section \ref{sec:results}, we show the results of our analysis and the cosmological inferences we can make. In section \ref{sec:desi_predictions}, we make some predictions for the precision at which the turnover can be measured using QSO observed as part of the Dark Energy Spectroscopic Instrument \citep[DESI;][]{DESI:2022xcl} survey as well as using high redshift probes from the Maunakea Spectroscopic Explorer \citep[MSE;][]{MSEScienceTeam:2019bva} and MegaMapper \citep{Schlegel:2019eqc}. Finally, in section \ref{sec:conclusions}, we outline our conclusions.

\section{Theory}
\label{sec:theory}

\subsection{The horizon scale at matter-radiation equality}

The epoch of matter-radiation equality is a major changing point in the history of the Universe. This is not just due to the change in dynamics, but also the change in the growth rate and propagation of the density perturbations. As the expansion decelerates, the Hubble-Lema\^itre rate falls, and super-horizon perturbations (that were seeded during cosmological inflation) can re-enter the horizon and begin to evolve. However, if they re-enter before matter-radiation equality, the pressure of the relativistic mass-energy that dominates provides a Jeans length equal to the horizon size, preventing gravitational collapse. It is only after the equality epoch that density perturbations can grow. The specific horizon size at the epoch matter-radiation equality is a significant scale in the structure formation history of the Universe, and one that depends only on the expansion history up to that point.

 The value of the horizon at the epoch of matter-radiation equality is given by
\begin{equation}
    r_\mathrm{H} = c\int_0^{a_\mathrm{eq}}\frac{\de a}{a^2 H(a)},
    \label{eq:rH}
\end{equation}
where the scale factor at matter-radiation equality $a_\mathrm{eq} = (1 + z_\mathrm{eq})^{-1} = \Omega_\mathrm{r}/\Omega_\mathrm{m}$ can be expressed in terms of the ratio of the radiation and matter density parameters. Since for $a < a_\mathrm{eq}$ the Universe was filled almost entirely with matter and radiation, we have $H(a) = H_0\sqrt{\Omega_\mathrm{m}}\sqrt{a_\mathrm{eq}/a^4 + 1/a^3}$ and, thus, the horizon scale
\begin{equation}
    r_\mathrm{H} = \frac{2c\left(\sqrt{2} - 1\right)\sqrt{a_\mathrm{eq}}}{H_0\sqrt{\Omega_\mathrm{m}}}
\end{equation}
is inversely proportional to the square-root of $\Omega_\mathrm{m}h^2$, for a fixed value of $a_\mathrm{eq}$. 

The horizon scale $r_\mathrm{H}$ can be related to the equality wavenumber $k_\mathrm{eq} = \left(4 - 2\sqrt{2}\right)r_\mathrm{H}^{-1}$ \citep{Prada:2011uz}. $k_\mathrm{eq}$ is approximately but not exactly the turnover scale $k_\mathrm{TO}$. $k_\mathrm{TO}$ has a weak dependence on the baryon density $\omega_\mathrm{b}$. While \citet{Prada:2011uz} provide a fitting formula
\begin{equation}
    k_\mathrm{TO} = \frac{0.194}{\omega_\mathrm{b}^{0.321}}k_\mathrm{eq}^{0.685 - 0.121\log_{10}\left(\omega_\mathrm{b}\right)}
    \label{eq:kmax_vs_keq}
\end{equation}
to connect the two scales, we are going to take a more accurate approach and find the maximum of fiducial power spectra computed using the Boltzmann solvers \textsc{camb} \citep*{Lewis:1999bs,Howlett:2012mh} and \textsc{class} \citep{Blas:2011rf}.

\subsection{Model independent parameterisation}

\begin{figure}
    \centering
    \includegraphics[width=\columnwidth]{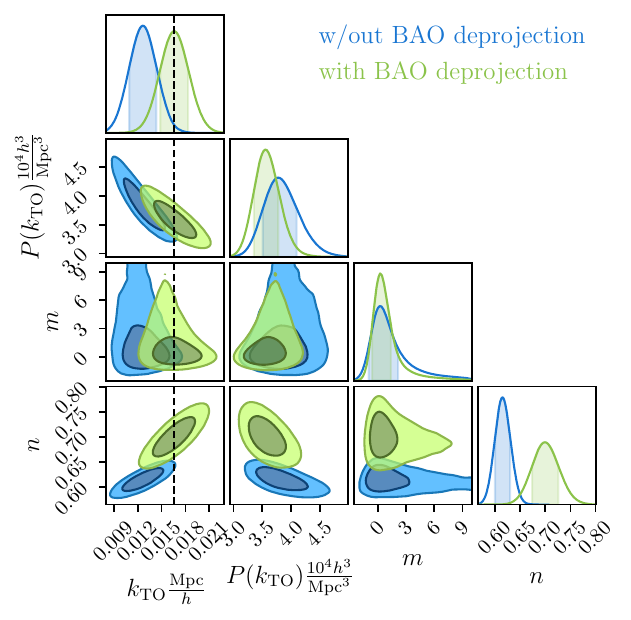}
    \caption{Parameter constraints from mock eBOSS QSO data obtained with and without BAO deprojection. The dashed line marks the input value of the turnover scale $k_\mathrm{TO}$.}
    \label{fig:eBOSS_sim_deproj_test}
\end{figure}

Following \citet{Poole:2012ex} and earlier work by \citet{Blake:2004tr}, we model the power spectrum as
\begin{equation}
    P(k) = \begin{cases} P_\mathrm{TO}^{1 - m x^2} & k < k_\mathrm{TO} \\
        P_\mathrm{TO}^{1 - n x^2} & k \geq k_\mathrm{TO},
    \end{cases}
    \label{eq:Poole_Pk}
\end{equation}
with $x = \log_{k_\mathrm{TO}\;\mathrm{Mpc}/h}\left(k\;\mathrm{Mpc}/h\right) - 1$. Here $P_\mathrm{TO}$, $k_\mathrm{TO}$, $m$ and $n$ are all free parameters that can be fit simultaneously using the data. The advantage of using this simple parameterisation is that we do not have to assume anything about the underlying cosmology and that we have a direct test of whether we have found the turnover, i.e. by estimating the fraction of the posterior volume with $m > 0$. On the other hand, we expect this model to describe the power spectrum well only close to the turnover. In particular, baryon acoustic oscillations (BAO) cause deviations from the asymmetric logarithmic hyperbola of \autoref{eq:Poole_Pk}. We demonstrate the impact of the BAO by generating mock power spectra distributed as described in \autoref{sec:fitting}. 
We then fit \autoref{eq:Poole_Pk} to the mock data and show the results from one typical realisation in blue in \autoref{fig:eBOSS_sim_deproj_test} where we can see that $k_\mathrm{TO}$ is degenerate with the slope $n$.
We also note that both parameters are biased with respect to the true input value. This is because the number of modes increases towards smaller scales, and, in turn, $n$ is very well constrained, but mostly from scales strongly affected by the BAO, and the slope decreases towards the smallest ones we consider. This is in line with \citet{Cunnington:2022ryj}, who found values of $k_\mathrm{TO}$ that are biased towards too low values in simulations of future line-intensity mapping surveys using the same power spectrum parameterisation. Therefore, he recommends using a logarithmic Laurent series parameterisation instead. However, such a parameterisation does not provide a direct test of whether the turnover has been detected or not. We have also explored alternative power spectrum parameterisations (cf. Appendix \ref{app:altparam}) but, after deprojecting the BAO as described as follows, we have found the parameterisation given in \autoref{eq:Poole_Pk} the most stable.

In lieu of changing the power spectrum parameterisation, we make the somewhat unconventional choice of treating the BAO signal as a systematic contaminant. 
We could overcome this systematic by properly modelling the BAO signal, but then we would become model-dependent, and our analysis would not be complementary to a full-shape analysis anymore. Instead, we trade precision for accuracy by down-weighting scales according to their contribution from the BAO. To do so, we first define a template
\begin{equation}
    \mathbf{f}^\mathrm{BAO}_k = \begin{cases} 0 & k < k_\mathrm{TO, fid} \\ P_\mathrm{fid}(k) - P_\mathrm{eq,BF}^{1 - n_\mathrm{BF} x^2} & k \geq k_\mathrm{TO, fid},\end{cases}
\end{equation}
by fitting \autoref{eq:Poole_Pk} to the power spectrum $P_\mathrm{fid}(k)$ computed by \textsc{class} for the fiducial cosmological parameters, fixing $k_\mathrm{TO}$ to its fiducial value $k_\mathrm{TO, fid}$ and varying $P_\mathrm{TO}$ and $n$ to find their best-fitting values $P_\mathrm{eq,BF}$ and $n_\mathrm{BF}$. We can then down-weight scales affected by the systematic described by the template $\mathbf{f}$ using \textit{mode deprojection}, first suggested by \citet{Rybicki:1992jz} and then adopted by \citet{Tegmark:1996qtQML}, \citet*{Slosar:2004fr}, \citet{Ho:2008bz}, \citet{Pullen:2012rd}, \citet{Leistedt:2013gfa}, \citet*{Leistedt:2014wia}, \citet*{Elsner:2015aga} and \citet{Kalus:2016cno} for various applications in cosmology. {The idea behind mode deprojection is to assign infinite variance to modes we believe are affected by a systematic contaminant, such as, here, the BAO. We, thus, model the power spectrum with BAO as $P_\mathrm{BAO}(k) = P(k) + \sqrt{\tau} \mathbf{f}^\mathrm{BAO}_k$, where $P(k)$ is the model power spectrum defined in \autoref{eq:Poole_Pk} and $\tau$ is the unknown amplitude parameter that accounts for the true power spectrum differing from $P_\mathrm{fid}(k)$.
Assuming $\mathbf{f}^\mathrm{BAO}_k$ to be independent of the BAO-free broad-band power spectrum $P(k)$, the covariance matrix reads}
\begin{equation}
    \tilde{\mathbfss{C}} = \mathbfss{C} + \lim_{\tau\rightarrow \infty}\tau\mathbf{f}^\mathrm{BAO}\mathbf{f}^\mathrm{BAO\dagger},
    \label{eq:modedeproj}
\end{equation}
where we marginalise over $\tau$ by taking the limit to infinity. With the Sherman-Morrison matrix inversion lemma
\begin{equation}
    \tilde{\mathbfss{C}}^{-1} = \mathbfss{C}^{-1} - \lim_{\tau\rightarrow \infty}\frac{\tau\mathbfss{C}^{-1}\mathbf{f}^\mathrm{BAO} \mathbf{f}^\mathrm{BAO\dagger}\mathbfss{C}^{-1}}{1 + \tau\mathbf{f}^\mathrm{BAO\dagger}\mathbfss{C}^{-1}\mathbf{f}^\mathrm{BAO}},
\end{equation}
the inverse of \autoref{eq:modedeproj} converges to
\begin{equation}
    \tilde{\mathbfss{C}}^{-1} = \mathbfss{C}^{-1} - \frac{\mathbfss{C}^{-1}\mathbf{f}^\mathrm{BAO} \mathbf{f}^\mathrm{BAO\dagger}\mathbfss{C}^{-1}}{\mathbf{f}^\mathrm{BAO\dagger}\mathbfss{C}^{-1}\mathbf{f}^\mathrm{BAO}}.
    \label{eq:moddep_inv_cov}
\end{equation}
We refit \autoref{eq:Poole_Pk} to the same mock power spectrum as before using the updated inverse covariance matrix given in \autoref{eq:moddep_inv_cov}, and we obtain unbiased results for all parameters including $k_\mathrm{TO}$ and $n$ (cf. green contours in \autoref{fig:eBOSS_sim_deproj_test}.

\begin{figure}
    \centering
    \includegraphics[width=\columnwidth]{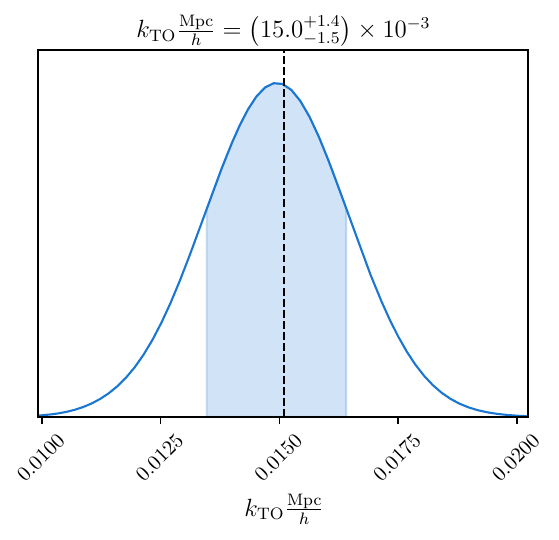}
    \caption{Marginalised $k_\mathrm{TO}$ posterior distribution from a mock power spectrum set up with $k_\mathrm{TO} = 0.0151h/\mathrm{Mpc}$ (marked with the dashed line), obtained with a mode deprojection template generated assuming $k_\mathrm{TO} = 0.0166h/\mathrm{Mpc}$.} 
    \label{fig:kmax_template_test}
\end{figure}

To test that our assumed value for $k_\mathrm{TO}$ in the template fit does not bias our $k_\mathrm{TO}$ constraints when $k_\mathrm{TO}$ differs from our assumption, we repeat the aforementioned check on a mock realisation where we set $\Omega_\mathrm{m} = 0.28$, corresponding to $k_\mathrm{TO} = 0.0151h/\mathrm{Mpc}$. As can be seen in \autoref{fig:kmax_template_test}, even in cases where the true $k_\mathrm{TO}$ and its value assumed in the template fitting differ, our results are consistent with the true turnover scale.

\subsection{Fitting}
\label{sec:fitting}

We use this BAO down-weighted covariance matrix when \textit{ensemble slice sampling} \citep{Karamanis:2020zss} $m$, $n$, $P_\mathrm{TO}$ and $k_\mathrm{TO}$ using the publicly available \textsc{zeus} code \citep*{Karamanis:2021tsx}. {Ensemble slice sampling is a black-box Markov Chain Monte Carlo algorithm that does not require the user to define any proposal distribution and has an acceptance rate of 1. The downside is that the posterior distribution function has to be evaluated more than once for each step. The algorithm requires an ensemble of random walkers who take each step simultaneously. Considering a walker at $x_n$, the next position $x_{n + 1}$ is determined by 
\begin{enumerate}
    \item drawing a random threshold $y_n$ uniformly distributed between zero and the value of the posterior distribution $\mathcal{P}$ evaluated at $x_n$,
    \item finding the interval for which $\mathcal{P}(x) > y_n$ along a slice through the parameter space. The slice goes through $x_n$, and its direction is defined by the unit vector between the position of two other random walkers chosen at random,
    \item $x_{n+1}$ is drawn uniformly from the slice interval.
\end{enumerate}} 

To be able to compare our model to data, we convolve it following \cite{Ross:2012sx}. {Considering a model power spectrum tabulated at wavenumbers $k_j$, we obtain the convolved model power spectrum at $k_i$ by the following quick matrix operation:
\begin{equation}
    P_\mathrm{conv}(k_i)=\sum_j \left[\left(W_{ij}-W_{0j}\frac{W(k_i)}{W(0)} \right)P(k_j)\right],
    \label{eq:Pconv}
\end{equation}}
{where the window matrix is defined as}
\begin{equation}
    W_{ij} = \int\de\varepsilon\int\de\cos(\theta) W(\varepsilon)\varepsilon^2 \Theta(r_\varepsilon, k_j ),
    \label{eq:window_matrix}
\end{equation}
{with $W(k_i)$ the window power spectrum which, in practice, we estimate from an unclustered random catalogue,} $r_\varepsilon=\sqrt{k_i^2+\varepsilon^2-2k_i\varepsilon\cos\theta}$, and $\Theta(r_\varepsilon, k_j )$ is one if $r_\varepsilon$ falls into the bin $k_j$ and zero otherwise. The second term in the round brackets {in \autoref{eq:window_matrix}} is the Fourier space analogue of the integral constraint on the correlation function.

Given a measurement $\tilde P(k)$ of the power spectrum, it is commonly assumed in the sampling process that the power spectrum is Gaussianly distributed, i.e.
\begin{equation}
    \chi^2 = \sum_{k_1 = k_\mathrm{min}}^{k_\mathrm{max}}\sum_{k_2 = k_\mathrm{min}}^{k_\mathrm{max}} \Delta P(k_1)\tilde{\mathbfss{C}}^{-1}_{k_1 k_2}\Delta P(k_2) + \text{const.}
    \label{eq:chi2}
\end{equation}
with $\Delta P(k) = P_\mathrm{conv}(k) - \tilde P(k)$ and $\tilde{\mathbfss{C}}_{k_1 k_2}$ the covariance matrix of $P_\mathrm{conv}(k)$ that, in practice, is evaluated for a specific fiducial cosmological model. This assumption is valid at scales where we have access to a huge enough number of $k$-modes to rely on the central limit theorem. However, current spectroscopic surveys only probe volumes that do not extend much further than the equality scale, and, in turn, constraints on $m$ come from only a small number of $k$-modes. 
In spite of this, these ultra-large scales are largely unaffected by non-linear clustering and the primordial density distribution is at least very close to being Gaussian \citep{Planck:nonGaussianity}. Assuming a perfectly Gaussian density field, its unconvolved power spectrum is Rayleigh distributed \citep*[e.g.][]{Kalus:2015lna}. However, since galaxies Poisson-sample the underlying (Gaussian) density field, one has to introduce scaling parameters $R$ and $\eta$ to the exponential that depend on the $k$-binning and the window convolution, resulting in a hypo-exponential distribution that can be approximated by a gamma distribution \citep{Wang:2018xuy}
\begin{equation}
    \mathcal{P}\left(\left.\tilde P(k)\right\vert R, \eta\right) = \frac{\eta^{-R}}{\Gamma(R)}\left[\tilde P(k)\right]^{R-1}\exp\left(-\frac{\tilde P(k)}{\eta}\right)\,,
    \label{eq:gamma_dist}
\end{equation}
with 
\begin{align}
    R=&\left[P_\mathrm{conv}(k)\right]^2/\left\langle\left[P_\mathrm{conv}(k) -\tilde  P(k)\right]^2\right\rangle,\nonumber\\
    \eta = &\left\langle\left[P_\mathrm{conv}(k) -\tilde  P(k)\right]^2\right\rangle/P_\mathrm{conv}(k)
\end{align} and $\Gamma(x)$ the extended factorial function. This can be Gaussianised by introducing a \citet{Box:1964}-transformed variable
\begin{equation}
    Z = \left[\tilde P(k)\right]^\nu, \; \nu > 0,
\end{equation}
where $\nu\approx 1/3$ is determined by demanding that the third central moment of the Gaussianised variable vanishes \citep{Wang:2018xuy}. With \autoref{eq:gamma_dist} and the Jacobian $\mathbfss{J}=\nu \left[\tilde P(k)\right]^{\nu - 1}$ one can compute the mean value 
\begin{equation}
    \left\langle Z \right\rangle = \frac{\Gamma\left(R + \nu\right)}{\Gamma(R)}\eta^\nu \approx R^\nu \eta^\nu = \left[P_\mathrm{conv}(k)\right]^\nu,
\end{equation}
where we have used the approximation $\Gamma\left(R + \nu\right)/\Gamma(R) \approx R^\nu$ \citep{Buric:2012} for $R\gg \nu$. We can thus modify \autoref{eq:chi2} by using the Gaussianised variable such that $\Delta P(k) = \left(\left\langle Z\right\rangle - Z\right)\mathbfss{J}^{-1}$ where the Jacobian $\mathbfss{J}$ comes from the rescaling of the covariance matrix:
\begin{equation}
    \Delta P(k) = 3\tilde P(k)\left(1 - \sqrt[3]{P_\mathrm{conv}(k)/\tilde P(k)}\right).
    \label{eq:Gaussianised_deltaP_from_gamma}
\end{equation}
This is similar to the inverse cubic normal distribution \citep[ICN;][]{Smith:2005ue,Kalus:2015lna} 
\begin{equation}
    \Delta P(k) = 3\tilde P(k)\left(1 - \sqrt[3]{\tilde P(k)/P_\mathrm{conv}(k)}\right)\,,
\end{equation}
which can be obtained by Taylor series matching around the peak of the power spectrum posterior resulting from multiplying single-mode Rayleigh distributions. As we are more interested in the overall shape of the posterior rather than only its peak, we are going to use the Gaussianised gamma distribution of \autoref{eq:Gaussianised_deltaP_from_gamma} in our inferences and compare it to inferences obtained with a standard (but wrong) Gaussian power spectrum distribution.

\section{Data}
\label{sec:data}

\begin{table}
    \centering
    \caption{Fiducial cosmological parameters assumed by \citet{Neveux:2020voa} to translate redshifts into distances, and hence throughout this paper.}
    \label{tab:fid_cosmo}    \begin{tabular}{l|c|c}
        \hline 
        Parameter & Symbol & Value/Relationship \\\hline
        Reduced Hubble-Lema\^itre constant & $h$ & $0.676$\\
        Hubble-Lema\^itre constant & $H_0$ & $100h\;\mathrm{km/s/Mpc}$\\
        Total matter abundance & $\Omega_\mathrm{m}$ & $0.31$\\
        Total radiation abundance & $\Omega_\mathrm{r}$ & $9.11\times 10^{-5}$ \\
        Reduced total matter abundance & $\omega_\mathrm{m}$ & $\Omega_\mathrm{m}h^2$\\
        Physical baryon density parameter & $\Omega_\mathrm{b}$ & $0.022/h^2$\\
        Cold dark matter density parameter & $\Omega_\mathrm{cdm}$ & $\Omega_\mathrm{m}-\Omega_\mathrm{b}$\\
        Dark energy density parameter & $\Omega_\Lambda$ & 
        $1-\Omega_\mathrm{m}$
        \\
        Amplitude of scalar fluctuations & $\sigma_8$ & $0.80$\\
        \hline
    \end{tabular}
\end{table}

To measure the turnover, we need a sample as large in volume as possible. The largest spectroscopically observed sample publicly available to date is the quasar catalogue of the final data release from the extended Baryon Oscillation Spectroscopic Survey (eBOSS). It comprises 343 708 quasars in the redshift range $0.8 < z < 2.2$, covering an effective area of $4699\, {\rm deg}^{2}$. eBOSS objects \citep{Ross:2020lqz} were selected from optical and infrared imaging data taken by the Sloan Digital Sky Survey \citep[SDSS]{SDSS:2000hjo} and the Wide Field Infrared Survey Explorer \citep[WISE]{Wright:2010qw}, respectively. Quasars are bright in the infrared, their target selection is thus less affected by angular variations in the imaging data caused, for instance, by foreground stars, and \citet{Mueller2} have successfully constrained primordial non-Gaussianity with eBOSS QSO data. On the other hand, these systematics are expected more pronounced in Emission Line (ELG) and Luminous Red Galaxy (LRG) samples, which have provided us with {outstanding} BAO measurements {\citep{Bautista:2020ahg,Gil-Marin:2020bct,Raichoor:2020vio,deMattia:2020fkb}} but have failed so far to produce cosmologically meaningful results at ultra-large scales \citep[e.g.][]{Ross:2012sx,Kalus:2018qsy}. Quasar catalogues also have a smoother radial selection function, making the clustering signal less susceptible to systematic small redshift misidentifications due to the observer's motion, the so-called \textit{Kaiser rocket effect}, as illustrated by \citet{Bahr-Kalus:2021jvu}. We, therefore, are not going to make use of the eBOSS ELG and LRG catalogues for turnover measurements. The quasar sample is still not free of spurious fluctuations in the density field caused by spatial variations in the quality of the imaging data, but due to it being the best sample to probe ultra-large scales, the eBOSS collaboration has {devoted substantial efforts towards mitigating systematic effects} at these scales. 

The power spectra used here\footnote{\hyperlink{https://github.com/mehdirezaie/eBOSSDR16QSOE}{https://github.com/mehdirezaie/eBOSSDR16QSOE}} have been obtained with systematic weights estimated by the novel neural network-based approach from \citet{Rezaie:2021voi}. This method, unlike the standard weights, accounts for galactic dust extinction using two tracer maps and for the effect of stars using the Gaia catalogue. This more rigorous systematic treatment does not affect scales smaller or comparable to the BAO scale. Therefore, the results and methods presented in \citet{Neveux:2020voa} hold and lay the foundations of \citet{Rezaie:2021voi}. \citet{Neveux:2020voa} translate the measured redshift $z$ of each quasar into a distance assuming the cosmological parameters given in \autoref{tab:fid_cosmo}. We adopt the same fiducial model to avoid recomputing the power spectra and systematic weights. \citet{Hou:2020rse} report the effective redshift of the eBOSS quasar sample as $z_\mathrm{eff} = 1.48$. For the fiducial model given in \autoref{tab:fid_cosmo}, the power spectra generated using the Boltzmann solvers \textsc{camb} \citep{Lewis:1999bs,Howlett:2012mh} and \textsc{class} \citep{Blas:2011rf} peak at 
\begin{equation}
    k_\mathrm{TO, fid} = 0.0166h/\mathrm{Mpc}.
    \label{eq:keq}
\end{equation}
To compute $r_\mathrm{H}$ using \autoref{eq:rH}, we compute $z_\mathrm{eq}$ as the ratio of our fiducial mater and radiation densities, which gives us
\begin{equation}
    r_\mathrm{H,fid} = 113.1\;\mathrm{Mpc}.
\end{equation}

\begin{figure}
    \centering
    \includegraphics[width=\columnwidth]{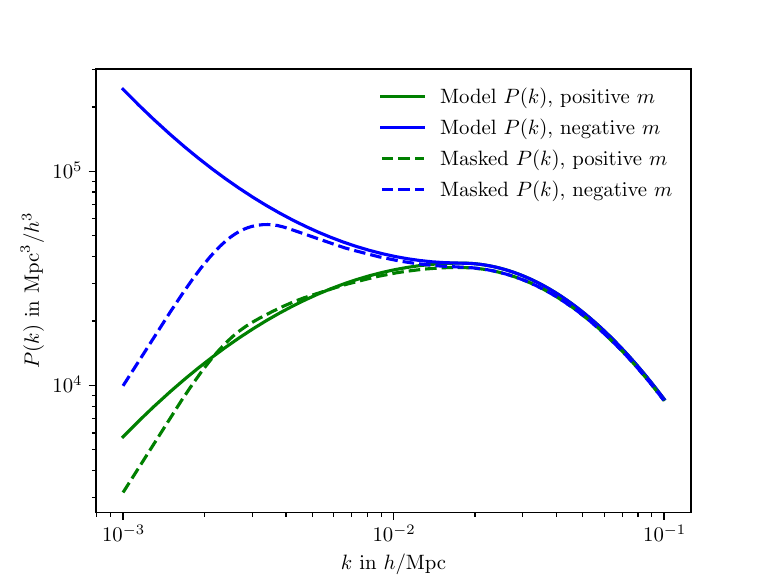}
    \caption{{The $P(k)$ model of \autoref{eq:Poole_Pk} evaluated for $k_\mathrm{TO} = 0.0177h/\mathrm{Mpc}$, $P_\mathrm{TO} = 37300\;\mathrm{Mpc}/h$, $m = \pm 0.35$ and $n= 0.755$. The dashed lines show the same power spectrum model convolved with the eBOSS QSO window function for the Northern hemisphere as described in \autoref{eq:Pconv}.}}
    \label{fig:window_effect}
\end{figure}
{In order to take the survey geometry into account, \citet{Rezaie:2021voi} have made use of a
catalogue of unclustered synthetic objects when computing their power spectra. This catalogue is often called the random
catalogue and matches the expected weighted density of
quasars, accounting for the radial and angular survey geometry. The details of generating the random catalogue are outlined in \citet{Ross:2020lqz}. We use the same random catalogues to compute the window power spectrum entering \autoref{eq:window_matrix}. We show the effect of the survey geometry on the power spectrum in \autoref{fig:window_effect} where we convolve model power spectra with the eBOSS QSO window function for the Northern hemisphere. For a positive value of $m$, the survey window introduces a slight flattening of the power spectrum peak and a sharp drop in power below $k \lesssim 0.002h/\mathrm{Mpc}$. For a negative value of $m$, corresponding to a power spectrum without a turnover, we see, as expected, an increase in power left to the turnover scale, but then also a sharp drop below $k \lesssim 0.002h/\mathrm{Mpc}$ similar to the case for positive $m$. This is the effect of the integral constraint, which stems from the fact that we do not know the Universe's average density and, therefore, assume that it equals the average density of the survey volume, making it hard to distinguish positive and negative $m$ with noisy data.}

\subsection{Testing the pipeline on mock data}

{\citet{Rezaie:2021voi} used synthetic catalogues called \textit{EZmocks} \citep{ZhaoMocks} to construct covariance matrices and perform robustness tests to characterise the significance of residual systematic uncertainties. The mocks were created using the extended \citet{ZeldovichApprox} approximation \citep{ChuangEZmock} and tuned to reproduce two-point clustering statistics of galaxies and quasars accurately. The EZmocks are manipulated to simulate the effects of observational systematics, including fibre collision, stellar contamination, redshift failures, and angular imaging systematics. The simulations are based on a flat $\Lambda$CDM cosmology with $h = 0.678$, $\Omega_\mathrm{m} = 0.307$, $\Omega_\mathrm{b} = 0.0482$, $\sigma_8 = 0.822$, and $n_\mathrm{s} = 0.961$. For these cosmological parameters, we expect a slightly lower value of turnover scale than for the fiducial eBOSS cosmology: $k_\mathrm{TO, EZmocks} = 0.0164h/\mathrm{Mpc}$. The mocks are constructed by combining periodic boxes with the comoving side length of $5\; \mathrm{Gpc}/h$ and subsampled along the line of sight to simulate the redshift distributions of the quasar sample. There are two sets of simulations, one with observational systematics and the other without. The covariance matrices are estimated from the latter, while the former is used to validate the eBOSS pipelines to mitigate the simulated systematic effects and recover the ground truth clustering.}

\begin{figure}
    \centering
    \includegraphics[width = \columnwidth]{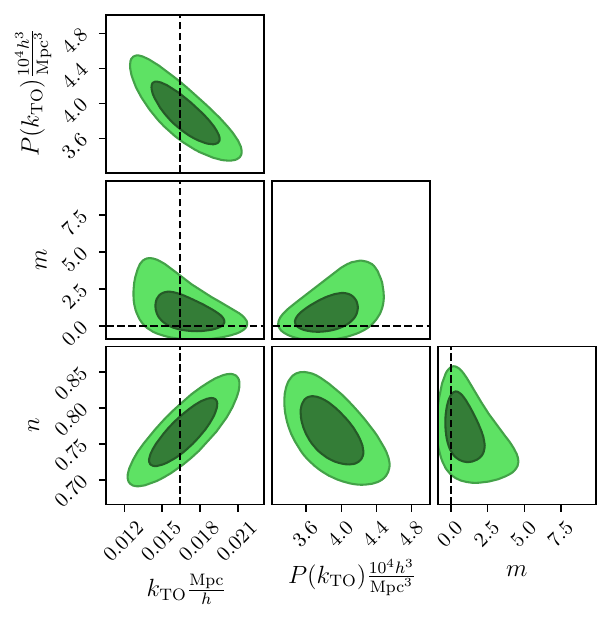}
    \caption{{Posterior contours of the turnover scale $k_\mathrm{TO}$, the power spectrum amplitude $P_\mathrm{TO}$ at that scale and the slope parameters $m$ and $n$ from the mean power spectra of 999 EZmocks. The contours have been computed under the assumption of a $\Gamma$-distributed power spectrum. The dashed lines mark the fiducial turnover scale $k_\mathrm{TO, EZmocks} = 0.0164h/\mathrm{Mpc}$.}}
    \label{fig:mock_mean}
\end{figure}

{We test our analysis pipeline on the uncontaminated null mocks. First, we run an MCMC analysis on the mean power spectra of the mocks. Unlike a single power spectrum, the Central Limit Theorem ensures that the mean of 999 power spectra is Gaussian distributed. We, therefore, only present our results from the Gaussianised $\Gamma$ distribution in \autoref{fig:mock_mean}; the posterior contours assuming a Gaussian likelihood is almost indistinguishable from the former. Our parameter constraints ($k_\mathrm{TO}=\left( 16.7^{+1.9}_{-1.7} \right) \times 10^{-3}h/\mathrm{Mpc}$, $P\left(k_\mathrm{TO}\right)=\left( 38.6^{+2.4}_{-2.3} \right) \times 10^{3}\mathrm{Mpc}^3/h^3$, $m = 0.58^{+1.04}_{-0.71}$ and $n = 0.764^{+0.032}_{-0.030}$) are consistent with our expectation of $k_\mathrm{TO, EZmocks} = 0.0164h/\mathrm{Mpc}$ and the critical value of $m = 0$ above which we have a turnover detection and translate into a 92 per cent probability of $m > 0$. }

\begin{figure}
    \centering
    \includegraphics[width=\columnwidth]{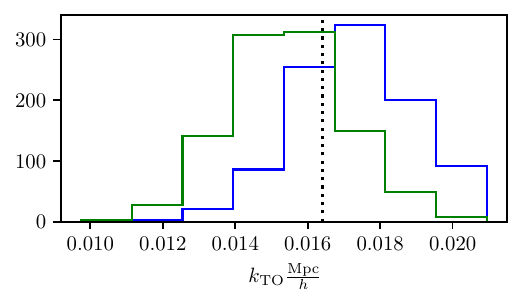}
    \includegraphics[width=\columnwidth]{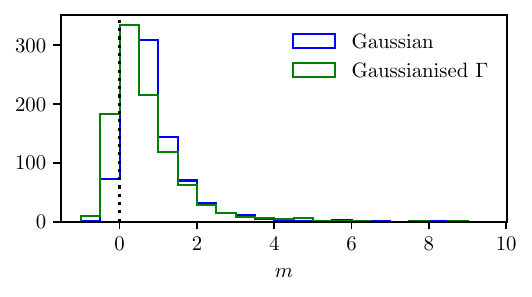}
    \caption{{Histograms of the best fitting $k_\mathrm{TO}$ and $m$ values from 999 EZmock realisations of an eBOSS QSO-like survey. The dashed line in the top panel marks the expected turnover scale $k_\mathrm{TO, EZmocks}$ in the cosmological model assumed in the generation of the mocks. The one in the bottom panel denotes $m=0$.}}
    \label{fig:mock_histoes}
\end{figure}

{We continue to find the best-fitting parameters of each mock realisation. To save computational resources, we avoid running MCMC by minimising the negative log-likelihood $\chi^2$ by applying the Broyden–Fletcher–Goldfarb–Shanno (BFGS) algorithm implemented in \textsc{scipy.optimize} \citep{scipy} with eight randomly chosen starting points to reduce the probability of finding local minima only. We show the histograms of the best-fitting values of $k_\mathrm{TO}$ and $m$ in \autoref{fig:mock_histoes}.When using a Gaussian likelihood, we find that 93 per cent of the EZmock power spectra have a positive best-fitting $m$, which is consistent with the results obtained from the MCMC analysis of the mean power spectra. However, the $k_\mathrm{TO}$ histogram peaks at values larger than the input value $k_\mathrm{TO,EZmocks} = 0.0164h/\mathrm{Mpc}$. Regardless, the input value lies at the 29-percentile, thus, within 1$\sigma$ of the median $k_\mathrm{TO}$ measurement assuming a Gaussian power spectrum distribution. }

\subsection{Testing the weights}

{When assuming the power spectra to be $\Gamma$-distributed, the fraction of mocks with positive best-fitting $m$ reduces to 81 per cent. On the other hand, the input value $k_\mathrm{TO, EZmocks}$ lies within the bin with the most values measured from the mocks.}

\begin{figure}
    \centering
    \includegraphics[width=\columnwidth]{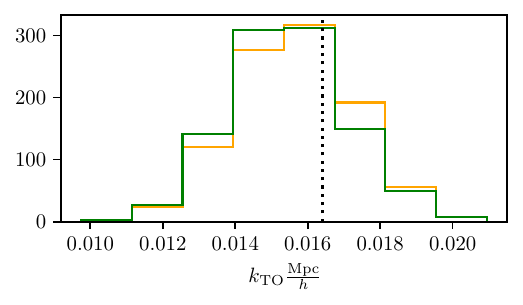}
    \includegraphics[width=\columnwidth]{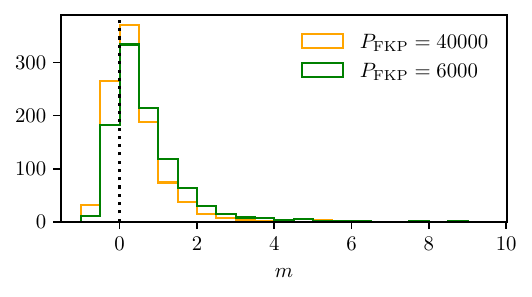}
    \caption{{Similar to \autoref{fig:mock_histoes} but comparing mock realisations with different FKP weights.}}
    \label{fig:PFKPalt_histoes}
\end{figure}

{Note that \citet{Rezaie:2021voi} applied \citet*[FKP]{FKP} weights
\begin{equation}
    w_\mathrm{FKP} = \frac{1}{1 + \bar n(z) P_\mathrm{FKP}}
\end{equation}
to each quasar in the eBOSS catalogue when computing the power spectra and covariance matrices, where $\bar n(z)$ is the average number density of quasars at redshift $z$ and $P_\mathrm{FKP}$ is the expected power spectrum at the scale for which the measurement shall be optimised. The standard value of $P_\mathrm{FKP}$ throughout eBOSS publications is $P_\mathrm{FKP} = 6000\;\mathrm{Mpc}^3/h^3$ to provide optimal BAO constraints. Increasing the value of $P_\mathrm{FKP}$ down-weights objects in high-density regions, effectively trading a decrease in cosmic variance for increased shot noise. As the power spectrum has a much larger amplitude at the turnover than at the BAO scale, we repeat the previous analysis with $P_\mathrm{FKP} = 40000\;\mathrm{Mpc}^3/h^3$, however, only using the Gaussianised $\Gamma$ distribution. We show again histograms of the resulting best-fitting values for $k_\mathrm{TO}$ and $m$ in \autoref{fig:PFKPalt_histoes}. We observe that the $k_\mathrm{TO}$ histogram becomes more symmetric with the increased $P_\mathrm{FKP}$. Still, when computing the standard deviation in $k_\mathrm{TO}$, we obtain a value of 0.0016 irrespective of the value for $P_\mathrm{FKP}$. This is in line with \citet{Mueller1}, where increasing $P_\mathrm{FKP}$ did not lead to improved constraints on the local primordial non-Gaussianity parameter $f_\mathrm{NL}$, possibly due to the overall sparsity of the quasar catalogue. We do see, however, an effect on the distribution of best-fitting $m$ parameters: Only 70 per cent of the $P_\mathrm{FKP} = 40000\;\mathrm{Mpc}^3/h^3$-mocks have a positive slope parameter $m$, thus, we expect a diminished detection probability when increasing $P_\mathrm{FKP}$ as we have degraded the power spectrum measurement also at scales larger as the turnover scale which are as important in measuring the slope as those close to the maximum. Due to this decrease in detection probability, we decide against modifying the FKP-weights.}

\section{Results}
\label{sec:results}

\subsection{Measurement of the turnover scale}

\begin{figure}
	\includegraphics[width=\columnwidth]{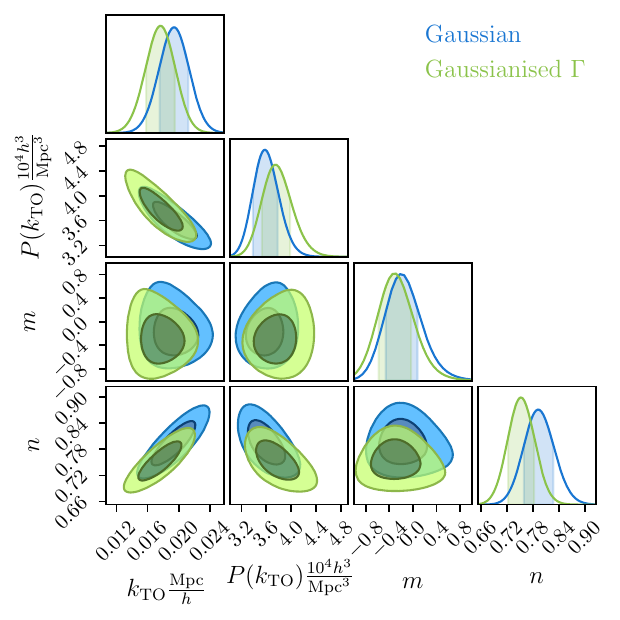}
    \caption{Posterior contours of the turnover scale $k_\mathrm{TO}$, the power spectrum amplitude $P_\mathrm{TO}$ at that scale and the slope parameters $m$ and $n$ {from eBOSS QSO data}. The blue contours have been obtained assuming a Gaussian likelihood, whereas the green ones have been computed under the more realistic assumption of a $\Gamma$-distributed power spectrum.}
    \label{fig:turnover_mcmc}
\end{figure}

\begin{figure*}
    \includegraphics[width=0.45\textwidth]{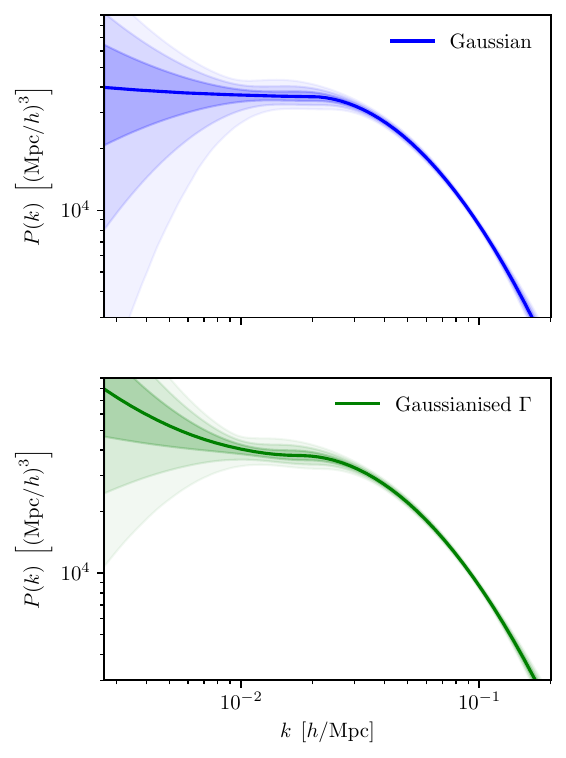}
    \includegraphics[width=0.45\textwidth]{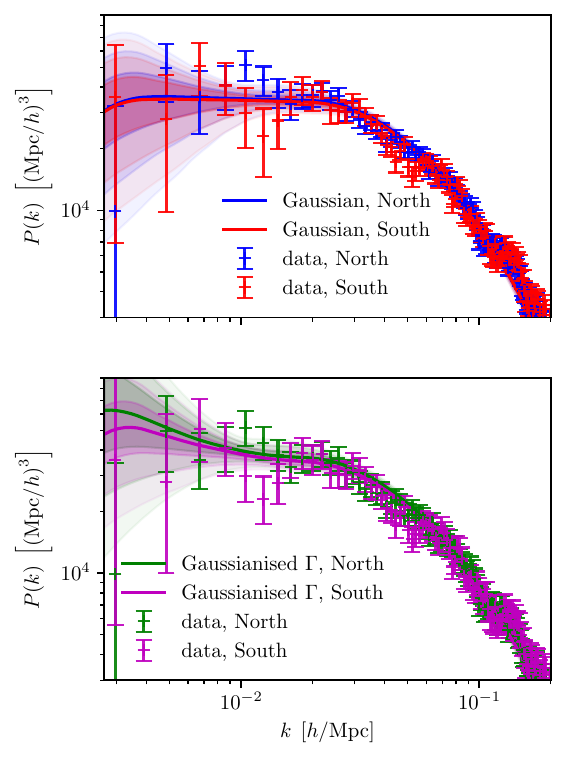}
    \caption{The median and 68, 95 and 99.5 percentiles of the model power spectra sampled from the MCMC chains used to obtain \autoref{tab:eBOSS_mcmc_results}. The right-hand (left-hand) panels show the (un)convolved power spectra. On the right, we also show the power spectra measured by the eBOSS collaboration \citep{Rezaie:2021voi} as crosses. The top (bottom) power spectra have been fitted, assuming a Gaussian (Gamma) distribution. {The error bars on the top right are the square root of the diagonal elements of the covariance matrix provided by \citet{Rezaie:2021voi}, whereas in the bottom right, we show errorbars corresponding to a Gamma distributed power spectrum (cf. \autoref{eq:gamma_dist}).}}
    \label{fig:Pk}
    \label{fig:Pk_masked}
\end{figure*}

We show in \autoref{fig:turnover_mcmc} the results from the two MCMC runs on eBOSS QSO data, assuming a Gaussian and a $\Gamma$-distributed power spectrum. We see in both cases that the data is consistent with $m = 0$, meaning that we have no evidence for a detection of the turnover scale. We reconstruct the power spectra for each set of parameters $\left\lbrace k_\mathrm{TO}, P_\mathrm{TO}, m, n\right\rbrace$ sampled by our MCMC chains and show their medians and 68, 95 and 99.5 percentiles in \autoref{fig:Pk}. We can clearly see an inflexion point in both panels, where the median power spectrum remains almost constant below the inflexion point in the Gaussian case, and it increases again when sampled from the $\Gamma$-distributed power. This seems counter-intuitive, but, under the assumption of a Gaussian density field, the power spectrum of a single mode is a squared quantity of a complex zero-centred Gaussian random variable. 
Negative values are hence impossible, and so, even if zero is the most likely value of the overdensity field, any deviation from zero in either the real or imaginary part leads to a positive value of the power spectrum, making a zero (or small) power spectrum very unlikely. Large power spectrum values can be realised by many combinations of real and imaginary overdensity values, causing the actual power spectrum errors to be skewed towards higher values compared to (falsely) assumed Gaussianity in the power spectrum. Better power spectrum data on large scales is therefore needed to find decisive evidence for the turnover.

Comparing the left- and right-hand panels of \autoref{fig:Pk}, we can remark yet another effect that diminishes the evidence for the turnover: due to the integral constraint in \autoref{eq:Pconv}, a turnover occurs naturally in the window-convolved power spectrum even for unconvolved power spectra with $m < 0$. A mildly negative $m$ can, therefore, fit some of the scatter close to the expected turnover scale and then can use the integral constraint to fit the lowest $k$ values.

\begin{table*}
    \centering
    \caption{Maximum posterior values of our eBOSS measurement assuming a Gaussian or {Gaussianised $\Gamma$} distribution for the power spectrum.}
    \label{tab:eBOSS_mcmc_results}
    \begin{tabular}{lcccccc}
        \hline
		Likelihood & $k_\mathrm{TO}$ [$h/\mathrm{Mpc}$] & $P_\mathrm{TO}$ [$\mathrm{Mpc}^3/h^3$] & $m$ & $n$ & $\alpha_\mathrm{TO}$ & $\alpha_\mathrm{eq}$\\ 
		\hline
            Gaussian & $\left( 19.4^{+1.9}_{-1.8} \right) \times 10^{-3}$ & $\left( 35.7^{+2.1}_{-1.9} \right) \times 10^{3}$ & $-0.23^{+0.31}_{-0.23}$ & $0.792^{+0.035}_{-0.033}$ & $1.17\pm 0.11$ & $1.19\pm 0.13$ \\ 
		Gaussianised $\Gamma$ & $\left( 17.6^{+1.9}_{-1.8} \right) \times 10^{-3}$ & $\left( 37.6\pm 2.3 \right) \times 10^{3}$ & $-0.32^{+0.29}_{-0.26}$ & $0.751^{+0.031}_{-0.028}$ & $1.06\pm 0.11$ & $1.07^{+0.12}_{-0.13}$ \\ 
		\hline
    \end{tabular}
\end{table*}

In the following, we assume that the inflexion point we found is indeed the turnover at the matter-radiation equality scale. Under this assumption, we measure $k_\mathrm{TO} = \left( 19.4^{+1.9}_{-1.8} \right) \times 10^{-3}h/\mathrm{Mpc}$ from the standard Gaussian power spectrum likelihood fit, thus, a value $1.4\sigma$ higher than our expectation for the fiducial cosmology (cf. \autoref{tab:eBOSS_mcmc_results}). However, using the $\Gamma$ likelihood, we find $k_\mathrm{TO} = \left( 17.6^{+1.9}_{-1.8} \right) \times 10^{-3}h/\mathrm{Mpc}$, i.e. a value consistent with the fiducial. This underlines the importance of accounting for the non-Gaussian distribution of the power spectrum at ultra-large scales.

\subsection{The turnover scale as a standard ruler}
As mentioned before, the turnover scale provides us with a BAO-independent standard ruler that potentially brings forth new insights into the Hubble tension. In analogy to BAO analyses, following \citet{SDSS:2005xqv}, we define a dilation measure
\begin{equation}
    D_\mathrm{V}(z) = \sqrt[3]{(1+z)^2D_\mathrm{A}^2(z)\frac{cz}{H(z)}},
    \label{eq:DV_def}
\end{equation}
where 
\begin{equation}
    D_\mathrm{A}(z) = \frac{c}{(1 + z)}\int_0^z\frac{\de z^\prime}{H\left(z^\prime\right)}
\end{equation}
is the angular diameter distance.
For the fiducial model of \autoref{tab:fid_cosmo}, we obtain a fiducial dilation measure of 
\begin{equation}
    D_{\mathrm{V,fid}}(z_\mathrm{eff} = 1.48) = 3821\;\mathrm{Mpc} = 33.79r_\mathrm{H,fid}.
\end{equation}
Again in analogy to BAO measurements, we introduce a stretch factor 
\begin{equation}
    \alpha_\mathrm{eq} = {\frac{k_\mathrm{eq}}{k_\mathrm{eq,fid}}} = \frac{D_\mathrm{V}(z)}{D_{\mathrm{V,fid}}(z)}\frac{r_\mathrm{H, fid}}{r_\mathrm{H}},
    \label{eq:alphaeq}
\end{equation}
where $r_\mathrm{H}$ takes the role of $r_\mathrm{drag}$. \footnote{Please note that this version of \autoref{eq:alphaeq} is different from that in previous literature and the published version of this paper, and so some numbers below will differ from this previous version, but match the values in the erratum.} However, we are measuring $k_\mathrm{TO}$ rather than $k_\mathrm{eq}$ and we are sensitive to the scaling factor
\begin{equation}
    \alpha_\mathrm{TO} = {\frac{k_\mathrm{TO}}{k_\mathrm{TO,fid}}}.
\end{equation}
With \autoref{eq:kmax_vs_keq} and fixing $\omega_\mathrm{b} = \omega_\mathrm{b,fid}$, we can relate the two alphas as
\begin{equation}
     \alpha_\mathrm{TO} \approx \alpha_\mathrm{eq}^{0.685 - 0.121\log_{10}\left(\omega_\mathrm{b, fid}\right)}.
\end{equation}
Our measurement of $k_\mathrm{TO} = \left( 17.6^{+1.9}_{-1.8} \right) \times 10^{-3}h/\mathrm{Mpc}$ and $k_\mathrm{TO, fid} = 0.0166h/\mathrm{Mpc}$ translate into a measurement of the stretch factor of $\alpha_\mathrm{TO} = {1.06\pm 0.11}$ using the Gaussianised $\Gamma$-likelihood for the power spectrum. This, in turn, can be recast into
\begin{equation}
    \alpha_\mathrm{eq} = {1.07^{+0.12}_{-0.13}}
\end{equation}
and
\begin{equation}
    D_\mathrm{V}(z_\mathrm{eff} = 1.48) = \left({36.2^{+4.1}_{-4.4}}\right)r_\mathrm{H}.
    \label{eq:DV_rH_measurement}
\end{equation}
Assuming $r_\mathrm{H} = r_\mathrm{H,fid}$, this reads
\begin{equation}
    D_\mathrm{V}(z_\mathrm{eff} = 1.48) = \left({4090^{+460}_{-500}}\right)\;\mathrm{Mpc}.
    \label{eq:DV_rHfid}
\end{equation}

The consensus BAO-only measurement of $D_\mathrm{V}(z_\mathrm{eff} = 1.48)$ is given by \citet{Neveux:2020voa} as
\begin{equation}
    D_\mathrm{V}(z_\mathrm{eff} = 1.48) = (26.51 \pm 0.42)r_\mathrm{drag}.
    \label{eq:DV_Neveux}
\end{equation}
For our fiducial cosmology, we have $r_\mathrm{drag, fid} = 147.07\;\mathrm{Mpc}$. We insert this into \autoref{eq:DV_Neveux} and find agreement between the eBOSS BAO measurement
\begin{equation}
    D_\mathrm{V}(z_\mathrm{eff} = 1.48) = (3899 \pm 62)\;\mathrm{Mpc}
\end{equation}
and the distance measurement given in \autoref{eq:DV_rHfid}.

\subsection{Energy density of the relativistic Universe}
\label{sec:energy_dens_of_rel_univ}

\begin{figure}
    \centering
    \includegraphics[width = \columnwidth]{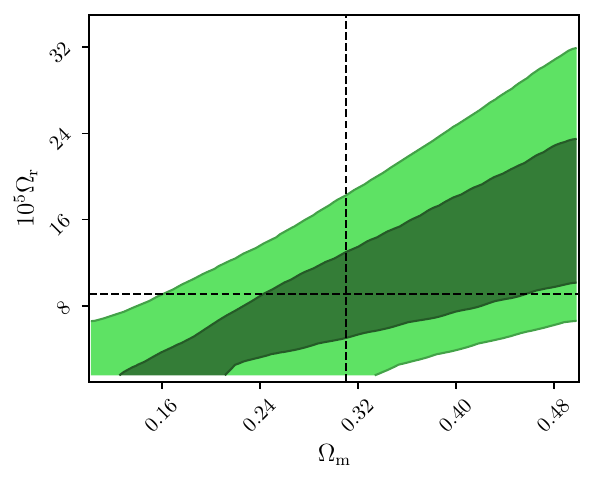}
    \caption{eBOSS QSO posterior contours in the $\Omega_\mathrm{m}$-$\Omega_\mathrm{r}$ plain. We explore $\Omega_\mathrm{m}$ only in the range $0.1 < \Omega_\mathrm{m} < 0.5$. The dashed lines correspond to the fiducial values of $\Omega_\mathrm{m} = 0.31$ and $10^5 \Omega_\mathrm{r} = 9.11$.}
    \label{fig:Om_Or}
\end{figure}

In \autoref{eq:DV_rH_measurement}, we have presented a model-independent, purely geometric measurement of $D_\mathrm{V}/r_\mathrm{H}$. As defined in \autoref{eq:DV_def}, $D_\mathrm{V}$ is an integrated quantity of $H(z)$ at low redshifts, whereas $r_\mathrm{H}$ depends on an integral over $H(z)$ from the end of inflation until matter-radiation equality (cf. \autoref{eq:rH}). Assuming a flat $\Lambda$CDM cosmology, $H(z)$ is determined by $H_0$ and $\Omega_\mathrm{m}$ at low redshifts and by $H_0$, $\Omega_\mathrm{r}$ and $\Omega_\mathrm{m}$ until matter-radiation equality. However, both $D_\mathrm{V}$ and $r_\mathrm{H}$ are proportional to $1/H_0$, making it, in principle, possible to constrain $\Omega_\mathrm{r}$ and $\Omega_\mathrm{m}$ from our eBOSS QSO turnover measurement alone without any external data. As can be seen in \autoref{fig:Om_Or}, $\Omega_\mathrm{r}$ and $\Omega_\mathrm{m}$ are highly degenerate, and we set a prior range of $0.1 < \Omega_\mathrm{m} < 0.5$ in order for our MCMC chains to converge. 

\begin{figure}
    \centering
    \includegraphics[width = \columnwidth]{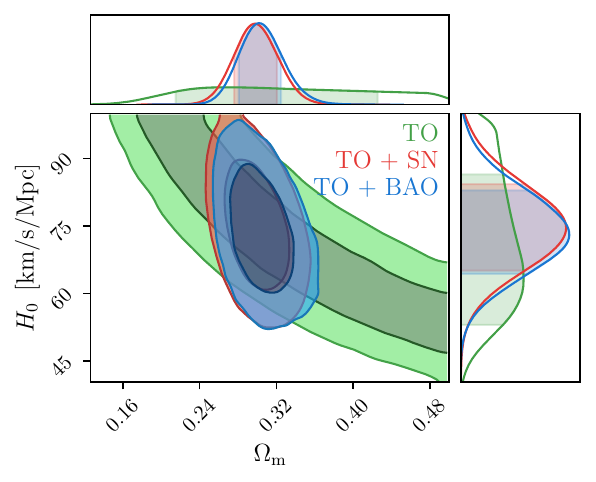}
    \includegraphics[width = \columnwidth]{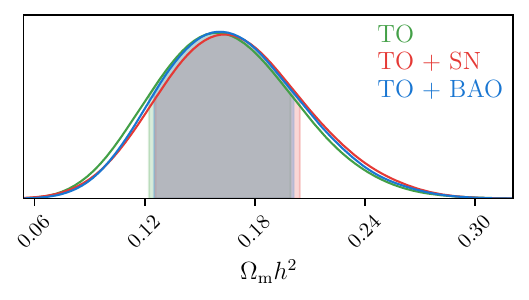}
    \caption{\textit{Top panel}: 2D constraints on the matter density $\Omega_m$ and the Hubble-Lema\^itre parameter $H_0$ from the eBOSS QSO turnover point (TO), in combination with Pantheon SN data as well as eBOSS BAO data. \textit{Bottom panel}: 1D posterior distribution of the combined parameter $\Omega_\mathrm{m}h^2$ from the same data combinations. }
    \label{fig:Om_H0_2D}
\end{figure}

Following the parameterisation of \citet{Komatsu}, the radiation density $\Omega_\mathrm{r}$ is related to the photon density $\Omega_\gamma$ as
\begin{equation}
    \label{eqn:omega_rel}
    \Omega_\mathrm{r} = \Omega_\gamma ( 1 + 0.2271 N_\mathrm{eff}),
\end{equation}
where $N_\mathrm{eff}$ is the effective number of extra relativistic degrees of freedom. Owing to the fact that neutrino decoupling was not instantaneous, its expected value is $N_\mathrm{eff} = 3.046$ \citep{Mangano}.
Assuming standard radiation physics, the physical photon energy density parameter is given by 
\begin{equation}
    \Omega_\gamma = \frac{8\pi G}{3 H_0^2}\frac{4\sigma_\mathrm{B}T_\mathrm{CMB}^4}{c^3},
\end{equation}
with $G$ denoting the gravitational constant, $\sigma_\mathrm{B}$ the Stefan-Boltzmann constant and $T_\mathrm{CMB}$ the mean CMB temperature. The latter is almost precisely known to be $T_\mathrm{CMB} = (2.72548 \pm 0.00057)\; \mathrm{K}$ \citep{Fixsen}. We fix $T_\mathrm{CMB}$, making $\Omega_\mathrm{r}$ effectively a function of $H_0$ only. Thus, we sample the $H_0$-$\Omega_\mathrm{m}$ posterior contour shown in {green} in {the bottom panel of} \autoref{fig:Om_H0_2D}, which is still degenerate and does not allow for competitive constraints on neither $H_0$ nor $\Omega_\mathrm{m}$ separately. However, {as shown in the bottom panel of \autoref{fig:Om_H0_2D},} we can resample this contour into the (almost) CMB-independent constraint of
\begin{equation}
    \Omega_\mathrm{m}h^2 = {0.159^{+0.041}_{-0.037}}.
\end{equation}
This is {high}, but consistent with $\Omega_\mathrm{m}h^2 = 0.1430 \pm 0.0011$ measured from Planck \citep{Planck:2018vyg}, providing a valuable consistency check.

{We also note that the turnover scale and the BAO scale are being measured roughly independently at the same redshift, and both measured quantities are a ratio with respect to the volumetric dilation measure $D_V(z)$. Therefore the ratio of the two is independent of this quantity, and measures the ratio of the horizon size of matter-radiation equality to the sound horizon at recombination. Combining our results with \autoref{eq:DV_Neveux} gives a value of this ratio of}
\begin{equation}
    \frac{r_d}{r_H} = {1.37^{+0.16}_{-0.17}} \,.
\end{equation}
{Comparing this to a fiducial value of $r_{d,\textrm{fid}}/r_{H,\textrm{fid}}=1.3$, we find we are still within 1-sigma of the expected value. This ratio could be used as a measurement of the expansion history between the two epochs, and so as a test of alternative cosmological models that suggest either some extra relativistic component, e.g dark radiation \citep[e.g.][]{riemer-sorensen_parkinson_davis_2013}, or some early dark energy \citep[e.g.][]{PhysRevLett.111.041301}. This measurement is independent of any CMB data, but not competitive with the constraints that are currently available in this sector from CMB datasets.}

\subsection{The inverse distance ladder and sound horizon-free constraints on the Hubble-Lema\^itre parameter}

The inverse distance ladder \citep{PhysRevD.92.123516,2015MNRAS.448.3463C,2019MNRAS.486.2184M} is a proposal that cosmological distance rulers that are anchored by physics in the early universe can be used to reconstruct or make inferences about the expansion history at low-redshift and Hubble-Lema\^itre constant at $z=0$. This is opposite to the usual cosmological distance ladder which builds up from local measurements of parallax and Cepheid variables to cosmological distances. Most inverse distance ladder inferences are calibrated using the drag scale $r_d$, which sets the comoving scale of the baryon acoustic oscillations. In this analysis, we use a different physical scale, the horizon size at matter-radiation equality, to make a probabilistic inference for the value of $H_0$.

While the measurement of $D_V(z)/r_\mathrm{H}$ can only constrain a combination of the Hubble-Lema\^itre parameter and the matter density, including additional constraints on the background evolution can break this degeneracy (see Fig.\ref{fig:Om_H0_2D}). (We also note that since the turnover scale is anchored to a scale set deep in the relativistic epoch, a single data point is also not sensitive to the value of the homogeneous curvature, as shown in Appendix \ref{app:curv}.)  We perform an MCMC analysis using \textsc{CosmoMC} \citep{Lewis:2002ah,Howlett:2012mh}, varying the Hubble-Lema\^itre parameter $H_0$, matter density $\Omega_m$ and baryon density $\Omega_b$ while assuming a flat $\Lambda$CDM cosmology. 
The CMB temperature and number of neutrino species are fixed, and so the relativistic energy density is given by Eq.~\ref{eqn:omega_rel}.  

Combining the equality scale measurement with the Pantheon supernova (SN) sample \citep{Pantheon} to constrain $\Omega_m$, we find a measurement  of the Hubble-Lema\^itre parameter of 
\begin{equation}
    H_0=\HOfromTOandSNeBF. 
\end{equation}
This measurement, unlike BAO constraints, is independent of the sound horizon and, furthermore, does not rely on the distance ladder calibration of the SN.
Similarly, we can include eBOSS LRG and Lyman-$\alpha$ BAO  \citep{eBOSS_DR16} constraints to include information on the matter density. Note that we do not include QSO BAO here to avoid possible correlations with our measurement of the turnover scale. The combination of the turnover measurement and the BAO measurement from eBOSS leads to an estimate of $H_0$ without including a prior on the baryon density from Big Bang nucleosynthesis (BBN). We find 
\begin{equation}
    H_0=\HOfromTOandBAOBF
\end{equation}
for the combination of eBOSS QSO turnover and BAO measurements. Both of our $H_0$ measurements are {high} compared to the \textit{Planck} measurement \citep{Planck:2018vyg} and {close to} the best fitting value of \citet{Riess:2021jrx}.

{While late-Universe probes are crucial to break the degeneracy between $H_0$ and $\Omega_\mathrm{m}$, we see in the bottom panel of \autoref{fig:Om_H0_2D} that they change our posterior distribution on $\Omega_\mathrm{m}h^2$ only marginally. This underlines the importance of $\Omega_\mathrm{m}h^2$ for constraining models of the early Universe.}

\section{Predictions for DESI, MSE and MegaMapper}
\label{sec:desi_predictions}
With the current data, we cannot report a decisive detection of the turnover scale. Here, we show that when DESI is completed, we expect to detect the turnover and that adding turnover measurements from the proposed Maunakea Spectroscopic Explorer (MSE) and MegaMapper that will extend spectroscopic clustering surveys to $z > 2$ will provide more constraining power on $H_0$ to help decide between the measurement of $H_0$ from either \citet{Planck:2018vyg} or \citet{Riess:2021jrx}.

\begin{table*}
    \centering
    \begin{tabular*}{\textwidth}{@{\extracolsep{\fill}}lccccc} \hline
         & eBOSS QSO & DESI QSO & MSE ELG & MSE LBG & MegaMapper\\ \hline
        $z_\mathrm{min}$ & 0.8 & 0.9 & 1.6 & 2.4 & 2.0\\
        $z_\mathrm{max}$ & 2.2 & 2.1 & 2.4 & 4.0 & 5.0\\
        $A$ & $4699\;\mathrm{deg}^2$ & $14000\;\mathrm{deg}^2$ & $10000\;\mathrm{deg}^2$ & $10000\;\mathrm{deg}^2$ & $14000\;\mathrm{deg}^2$ \\
        $r_{\parallel, \mathrm{max}}$ & $2725\;\mathrm{Mpc}$ & $2315\;\mathrm{Mpc}$ & $1189\;\mathrm{Mpc}$ & $1489\;\mathrm{Mpc}$ & $2645\;\mathrm{Mpc}$\\ 
        $r_{\perp, \mathrm{max}}$ & $6307\;\mathrm{Mpc}$ & $9380\;\mathrm{Mpc}$ & $8975\;\mathrm{Mpc}$ & $11258\;\mathrm{Mpc}$ & $13677\;\mathrm{Mpc}$\\
        $k_{\parallel, \mathrm{min}}$ & $0.0034h/\mathrm{Mpc}$ & $0.0040h/\mathrm{Mpc}$ & $0.0078h/\mathrm{Mpc}$ & $0.0062h/\mathrm{Mpc}$ & $0.0035h/\mathrm{Mpc}$\\
        $k_{\perp, \mathrm{min}}$ & $0.0015h/\mathrm{Mpc}$ & $0.00099h/\mathrm{Mpc}$ & $0.0010h/\mathrm{Mpc}$ & $0.00083h/\mathrm{Mpc}$ & $0.00068h/\mathrm{Mpc}$\\
        $\bar n$ & $4.83\times 10^{-6}/\mathrm{Mpc^3}$ & $1.50\times 10^{-5}/\mathrm{Mpc^3}$ & $5.56\times 10^{-5}/\mathrm{Mpc^3}$ & $3.40\times 10^{-5}/\mathrm{Mpc^3}$ & $7.88\times 10^{-5}/\mathrm{Mpc^3}$\\
        \hline
    \end{tabular*}
    \caption{Comparison of survey extents of the eBOSS and DESI quasar samples, the Maunakea Spectroscopic Explorer (MSE) and MegaMapper. The DESI QSO numbers are taken from \citet{Yeche:2020tjm}, the MSE specifications from table 9 of \citet{MSEScienceTeam:2019bva} and the MegaMapper figures from table 2 of \citet{Ferraro:2019uce}.}
    \label{tab:kmin_eBOSS_DESI}
\end{table*}

The Fisher information of the power spectrum $F\propto V_\mathrm{eff}$ is proportional to the effective volume
\begin{equation}
    V_\mathrm{eff}(k) = V_\mathrm{survey}\left(\frac{\bar n P(k)}{1 + \bar n P(k)}\right)^2, 
\end{equation}
with $V_\mathrm{survey}$ the survey volume and $\bar n$ the average number density within $V_\mathrm{survey}$. Due to this proportionality, a rough estimate of the DESI covariance is 
\begin{equation}
    \mathbfss{C}_\mathrm{DESI}^{k_1 k_2} = \sqrt{\frac{V_\mathrm{eff}^\mathrm{eBOSS}(k_1)}{V_\mathrm{eff}^\mathrm{DESI}(k_1)}}\mathbfss{C}_\mathrm{eBOSS}^{k_1 k_2}\sqrt{\frac{V_\mathrm{eff}^\mathrm{eBOSS}(k_2)}{V_\mathrm{eff}^\mathrm{DESI}(k_2)}}.
    \label{eq:covariance_rescaling}
\end{equation}
The DESI catalogue of quasars as direct tracers (as opposed to the catalogue of tracers used for measurements of the Lyman alpha forest) spans a slightly narrower redshift range than eBOSS, i.e. $0.9 < z < 2.1$ \citep[][cf. \autoref{tab:kmin_eBOSS_DESI}]{Yeche:2020tjm}. However, DESI will target about three times the eBOSS area. While eBOSS probes ultra-large scales mostly through radial modes, DESI will thus measure angular modes that are almost as long as the radial modes. We estimate the minimal attainable wave numbers of the two surveys in radial and angular directions as $k_{\parallel, \mathrm{min}} = 2\pi/r_{\parallel, \mathrm{max}}$ and $k_{\perp, \mathrm{min}} = 2\pi/r_{\perp, \mathrm{max}}$, where (approximating the survey footprints as circles)
\begin{align}
    r_{\parallel, \mathrm{max}} &= d(z_\mathrm{max}) - d(z_\mathrm{min}) \text{ and} \nonumber\\
    r_{\perp, \mathrm{max}} &= 2d(z_\mathrm{max})\sin\left(\frac{\sqrt{A}}{2}\right)
\end{align}
with $d(z)$ the comoving distance to redshift $z$ and $A$ the solid angle covered by the survey. Evaluating this for the survey specifications listed in \autoref{tab:kmin_eBOSS_DESI} and our fiducial cosmology, we find that DESI will not probe any new scales, instead, it is expected to reduce the uncertainty on the power spectrum measurement due to probing a larger volume. Furthermore, DESI is targeting 2.1 million tracer quasars, or in other words six times the number of eBOSS quasars. Thus, the final DESI sample will be about twice as dense as the eBOSS sample, increasing the effective volume by a factor of four (assuming $P\left(k_{TO}\right) = 37300\;\mathrm{Mpc}^3/h^3$ in line with our results in \autoref{tab:eBOSS_mcmc_results}).

\begin{figure}
    \centering
    \includegraphics[width = \columnwidth]{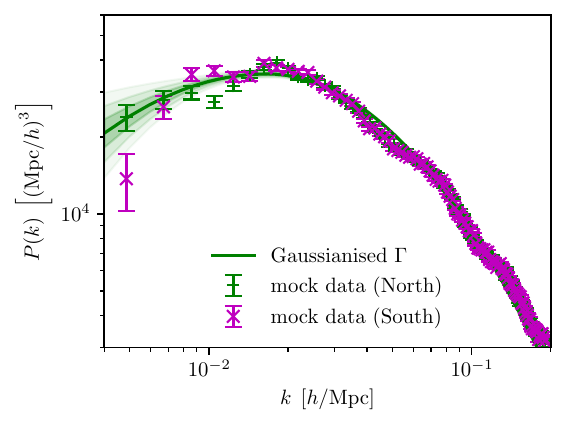}
    \caption{Similar to \autoref{fig:Pk}, but for a `mock power spectrum' generated using \textsc{class} and a rescaled covariance matrix predicting DESI errors. The error bars include the contribution from the BAO deprojection (cf. \autoref{eq:modedeproj}).}
    \label{fig:DESI_Pk}
\end{figure}

\begin{table*}
    \centering
    \caption{Maximum posterior values of our forecast{s}.}
    \label{tab:desi_forecast}
    \begin{tabular*}{\textwidth}{@{\extracolsep{\fill}}cccccccc}
        \hline
		Survey & $k_\mathrm{TO}$ [$h/\mathrm{Mpc}$] & $m$ & $n$ & $\mathcal{P}(m > 0)$ & $\alpha_\mathrm{TO}$ & $\alpha_\mathrm{eq}$\\ 
		\hline
		{DESI QSO} & $\left( 168.7^{+4.7}_{-4.3} \right) \times 10^{-4}$ & $0.22^{+0.13}_{-0.12}$ & $\left( 695.6^{+5.8}_{-6.2} \right) \times 10^{-3}$ & 0.97 & $1.016^{+0.028}_{-0.026}$ & $1.018^{+0.032}_{-0.029}$ \\ 
            {MSE ELG} & $\left( 170.5^{+3.4}_{-3.6} \right) \times 10^{-4}$ & $0.23^{+0.13}_{-0.12}$ & $\left( 698.1^{+4.1}_{-3.9} \right) \times 10^{-3}$ & 0.98 & $1.027^{+0.021}_{-0.022}$ & $1.031^{+0.023}_{-0.025}$\\
            {MSE LBG} & $\left( 168.5^{+3.0}_{-2.9} \right) \times 10^{-4}$ & $0.263^{+0.107}_{-0.095}$ & $\left( 697.4^{+3.5}_{-3.6} \right) \times 10^{-3}$ & 0.998 & $1.015^{+0.018}_{-0.017}$ & $1.017^{+0.021}_{-0.019}$\\
            {MegaMapper} &  $\left( 166.2\pm 1.5 \right) \times 10^{-4}$ & $0.384\pm 0.063$ & $\left( 697.8^{+1.7}_{-1.5} \right) \times 10^{-3}$ & 0.999997 & $\left( 1000.9\pm 8.8 \right) \times 10^{-3}$ & $\left( 1001.2^{+9.9}_{-10.0} \right) \times 10^{-3}$\\
		\hline
    \end{tabular*}
\end{table*}

We generate a mock power spectrum by running \textsc{class} for our fiducial cosmology. Fitting the mock power spectrum to \autoref{eq:Poole_Pk}, we detect the turnover with a {96} per cent chance (cf. \autoref{tab:desi_forecast}). Presuming a $\Gamma$-distributed
power spectrum, we predict $\alpha_\mathrm{eq} = 1.018^{+0.032}_{-0.029}$ for DESI. We again convert this into a distance measurement of $D_\mathrm{V}(z_\mathrm{eff} = 1.48) = \left({34.40_{-0.98}^{+1.08}}\right)r_\mathrm{H}$. Following the same routine as described in \autoref{sec:energy_dens_of_rel_univ}, this measurement translates into a constraint of 
\begin{equation}
    \Omega_\mathrm{m}h^2 = 0.135^{+0.030}_{-0.013}
\end{equation}
As future SN and BAO measurements will be independent of $H_0$ (cf. \autoref{fig:Om_H0_2D}), we can obtain an optimistic forecast on $H_0$ from combining DESI BAO and DESI QSO turnover measurements by slicing our $\Omega_\mathrm{m}$-$H_0$ contour at our fiducial value of $\Omega_\mathrm{m} = 0.31$:
\begin{equation}
    H_0 = \left(66.3^{+7.2}_{-2.9}\right)\;\mathrm{km/s/Mpc}.
\end{equation}
As expected, this is consistent with our fiducial $H_0$ value but {also} with \citet{Riess:2021jrx}'s local measurement of $H_0 = \left(73.04 \pm 1.04\right)\;\mathrm{km/s/Mpc}$.

\begin{figure}
    \centering
    \includegraphics[width = \columnwidth]{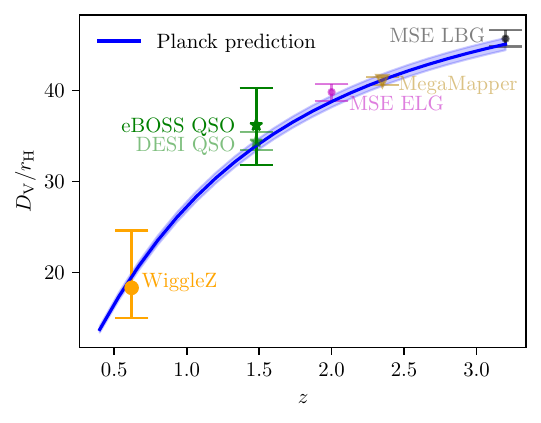}
    \caption{$D_\mathrm{V}/r_\mathrm{H}$ measurements from WiggleZ and eBOSS QSOs (this work) along with forecasts for DESI QSO, MSE and MegaMapper. The blue line represents the predicted curve from \textit{Planck} best fitting cosmological parameters along with its $3\sigma$ error contour in light blue. The offset of the forecasts compared to the \textit{Planck} prediction manifests the difference between \citet{Neveux:2020voa}'s fiducial matter density parameter, adopted in the forecasts, and the Planck measurement. }
    \label{fig:DV_over_rH_forecasts}
\end{figure}

We make similar forecasts to check whether we can improve on the turnover $H_0$-constraints with two planned surveys, namely the Maunakea Spectroscopic Explorer \citep[MSE;][]{MSEScienceTeam:2019bva} and MegaMapper \citep{Schlegel:2019eqc}. We also list the parameters that we have to assume for these surveys in \autoref{tab:kmin_eBOSS_DESI}. We rescale the eBOSS covariance matrix in the same way as in \autoref{eq:covariance_rescaling}. In principle, we would also have to rescale the covariance considering lower values of the growth factor and the galaxy bias. However, the same rescaling would be needed in the power spectrum, which then cancels out in the likelihood function. Though we consider these when computing the effective volume $V_\mathrm{eff}$. Probing considerably larger volumes than eBOSS and DESI, MSE and MegaMapper will have access to lower $k$-values. Yet, with the same $k$-binning as the one used in the eBOSS analysis, we can only accommodate one more $k$-bin at ultra-large scales. Lacking eBOSS covariances that we could rescale at these scales, we are conservative and do not consider additional large-scale 
information. We again fit \autoref{eq:Poole_Pk} to fiducial \textsc{CLASS} power spectra using the rescaled covariance matrices, translate our results into $D_\mathrm{V}(z_\mathrm{eff})/r_\mathrm{H}$-measurements, plot our results in \autoref{fig:DV_over_rH_forecasts} and list our forecast $\alpha_\mathrm{eq}$ values in \autoref{tab:future_forecast}. MegaMapper will provide an $\alpha_\mathrm{eq}$ constraint that is tighter than the isotropic $\alpha_\mathrm{bao}$ from eBOSS QSO \citep{Hou:2020rse}. We translate these forecasts into constraints on $\Omega_\mathrm{m}h^2$ with MegaMapper expected to yield $\Omega_\mathrm{m}h^2 = \left( 136.7^{+17.7}_{-4.6} \right) \times 10^{-3}$, a factor-of-{2} improvement on our eBOSS QSO measurement, but still an order of magnitude wider than current \textit{Planck} constraints. When combined with their BAO counterparts, MSE and MegaMapper {may} supply tight enough $H_0$ constraints that are independent of both SN and CMB measurements that allow us to favour 
either \citet{Planck:2018vyg} or \citet{Riess:2021jrx}'s $H_0$-measurement (see also \autoref{tab:future_forecast}).

\begin{table}
    \centering
    \caption{{Forecasted values of $\Omega_\mathrm{m}h^2$ from the TO alone and $H_0$ after combination with BAO or SN data to break the $\Omega_\mathrm{m}$-$H_0$ degeneracy. }}
    \label{tab:future_forecast}
    \begin{tabular*}{\columnwidth}{@{\extracolsep{\fill}}lcc}
        \hline
		Sample & $\Omega_\mathrm{m}h^2$ & $H_0$ [km/s/Mpc]\\ 
		\hline
		DESI QSO & $0.135^{+0.030}_{-0.013}$ & $66.3^{+7.2}_{-2.9}$ \\ 
		MSE ELG & $0.139^{+0.026}_{-0.010}$ & $67.0^{+6.3}_{-2.3}$ \\ 
		MSE LBG & $\left( 138.5^{+22.8}_{-7.7} \right) \times 10^{-3}$ & $66.9^{+5.4}_{-1.7}$ \\ 
		MegaMapper & $\left( 136.7^{+17.7}_{-4.6} \right) \times 10^{-3}$ & $66.4^{+4.2}_{-1.0}$ \\
    \hline
    \end{tabular*}
\end{table}

\section{Conclusions}
\label{sec:conclusions}

We have measured the power spectrum turnover induced at the epoch of matter-radiation equality in a model-independent fashion by fitting an asymmetric logarithmic hyperbola to the power spectrum measured from the quasar sample of the final data release of the extended Baryon Oscillation Spectroscopic Survey. This provides a standard ruler independent of the BAO and, hence, of the sound horizon at last scattering, providing a useful cross-check of BAO distance measures.

\begin{itemize}
    \item While the scale of the turnover can be detected at high significance, the eBOSS measurement of the power spectrum is not precise enough to determine its gradient on scales larger than the turnover that are set by $n_{\mathrm{s}}$, the spectral index of the primordial power spectrum.
    \item While we find the scale of the turnover in agreement with our expectation as long as we approximate the power spectrum likelihood by a $\Gamma$ distribution, the common assumption of a Gaussianly distributed power spectrum breaks down at ultra-large scales due to small number statistics and biases the turnover measurement towards smaller scales. This underlines the importance of using the correct posterior distribution when probing ultra-large scales.
    \item We resort to the turnover scale as a standard ruler in analogy to BAO measurements and find $D_\mathrm{V}(z_\mathrm{eff} = 1.48) = \left({36.2^{+4.1}_{-4.4}}\right)r_\mathrm{H}$.
    \item Assuming standard radiation physics and a flat $\Lambda$CDM cosmology, this geometric measurement translates into a cosmological measurement of $\Omega_\mathrm{m}h^2 = {0.159^{+0.041}_{-0.037}}$.
    \item Combining the above measurement with low-redshift distance measurements from Pantheon SNe or eBOSS BAO, we obtain a sound-horizon free estimate of the Hubble-Lema\^itre parameter $H_0$ that is also free from any rare astronomical distance ladder anchors. We find it to be $H_0=\HOfromTOandSNeBF$ in the former, and $H_0=\HOfromTOandBAOBF$ in the latter case. In either case, we are {close to} the best fitting value of \citet{Riess:2021jrx} {with a less than a $\sigma$ preference over the \citet{Planck:2018vyg} results}.
    \item With the increase in volume and number of quasars observed by DESI, we forecast that DESI will establish evidence for the turnover at {97} per cent confidence level. DESI will bring down the uncertainty on $\Omega_\mathrm{m}h^2$ to $\sigma\left(\Omega_\mathrm{m}h^2\right) = 0.021$ and on $H_0$ to $\sigma\left(H_0\right) = {5.1}\;\mathrm{km/s/Mpc}$. 
    \item Making forecasts for MSE (MegaMapper), we estimate $\sigma(H_0) = 3.6\;\mathrm{km/s/Mpc}$ (${2.6}\;\mathrm{km/s/Mpc}$), less than the tension between the Planck and SH0ES constraints on $H_0$.
\end{itemize}

\section*{Acknowledgements}

The authors express their gratitude to Mehdi Rezaie for making the eBOSS quasar catalogue, covariance matrices and power spectra with his increased systematic weights publicly available, and for providing us with the EZmocks used for his analysis. {They would also like to extend their heartfelt appreciation to Arnaud de Mattia, whose keen observation helped identify and rectify inconsistencies in earlier versions of this article.} They would like to thank Sungwook E. Hong, Cullan Howlett and Licia Verde for valuable discussions around the project. {The authors acknowledge the invaluable contributions of the anonymous referee, whose insightful comments and suggestions greatly improved the quality of this work.} BBK and DP are supported by the project \begin{CJK}{UTF8}{mj}우주거대구조를 이용한 암흑우주 연구\end{CJK} (``Understanding Dark Universe Using Large Scale Structure of the Universe''), funded by the Ministry of Science of the Republic of Korea. E-MM acknowledges support from the European Research Council (ERC) under the European Union’s Horizon 2020 research
and innovation programme (grant agreement No 693024). 

This work was supported by the high-performance computing clusters Seondeok at the Korea Astronomy and Space Science Institute. This research made substantial use of \textsc{camb} \citep{Lewis:1999bs,Howlett:2012mh}, \textsc{class} \citep{Blas:2011rf}, the \textsc{ChainConsumer} package \citep{samuel_hinton_2020_4280904}, \textsc{CosmoMC} \citep{Lewis:2002ah,Howlett:2012mh}, the \textsc{Numpy} package \citep{Numpy}, \textsc{Matplotlib} \citep{Hunter:2007}, {\textsc{SciPy} \citep{scipy}} and \textsc{zeus} \citep{Karamanis:2021tsx}.

\section*{Data Availability}

The eBOSS power spectrum data used in this analysis were generated elsewhere and can be found at \url{https://github.com/mehdirezaie/eBOSSDR16QSOE}. The MCMC chains generated as part of this analysis are available upon request to the authors.

\bibliographystyle{mnras}
\bibliography{equalityscalefromeBOSS}

\begin{thebibliography}{}
\makeatletter
\relax
\def\mn@urlcharsother{\let\do\@makeother \do\$\do\&\do\#\do\^\do\_\do\%\do\~}
\def\mn@doi{\begingroup\mn@urlcharsother \@ifnextchar [ {\mn@doi@}
  {\mn@doi@[]}}
\def\mn@doi@[#1]#2{\def\@tempa{#1}\ifx\@tempa\@empty \href
  {http://dx.doi.org/#2} {doi:#2}\else \href {http://dx.doi.org/#2} {#1}\fi
  \endgroup}
\def\mn@eprint#1#2{\mn@eprint@#1:#2::\@nil}
\def\mn@eprint@arXiv#1{\href {http://arxiv.org/abs/#1} {{\tt arXiv:#1}}}
\def\mn@eprint@dblp#1{\href {http://dblp.uni-trier.de/rec/bibtex/#1.xml}
  {dblp:#1}}
\def\mn@eprint@#1:#2:#3:#4\@nil{\def\@tempa {#1}\def\@tempb {#2}\def\@tempc
  {#3}\ifx \@tempc \@empty \let \@tempc \@tempb \let \@tempb \@tempa \fi \ifx
  \@tempb \@empty \def\@tempb {arXiv}\fi \@ifundefined
  {mn@eprint@\@tempb}{\@tempb:\@tempc}{\expandafter \expandafter \csname
  mn@eprint@\@tempb\endcsname \expandafter{\@tempc}}}

\bibitem[\protect\citeauthoryear{{Abareshi} et~al.,}{{Abareshi}
  et~al.}{2022}]{DESI:2022xcl}
{Abareshi} B.,  et~al., 2022, \mn@doi [\aj] {10.3847/1538-3881/ac882b}, \href
  {https://ui.adsabs.harvard.edu/abs/2022AJ....164..207A} {164, 207}

\bibitem[\protect\citeauthoryear{Aghanim et~al.}{Aghanim
  et~al.}{2020}]{Planck:2018vyg}
Aghanim N.,  et~al., 2020, \mn@doi [Astron. Astrophys.]
  {10.1051/0004-6361/201833910}, 641, A6

\bibitem[\protect\citeauthoryear{Akrami et~al.}{Akrami
  et~al.}{2020}]{Planck:nonGaussianity}
Akrami Y.,  et~al., 2020, \mn@doi [Astron. Astrophys.]
  {10.1051/0004-6361/201935891}, 641, A9

\bibitem[\protect\citeauthoryear{{Alam} et~al.,}{{Alam}
  et~al.}{2017}]{BOSS:2016wmc}
{Alam} S.,  et~al., 2017, \mn@doi [\mnras] {10.1093/mnras/stx721}, \href
  {https://ui.adsabs.harvard.edu/abs/2017MNRAS.470.2617A} {470, 2617}

\bibitem[\protect\citeauthoryear{Alam et~al.,}{Alam et~al.}{2021}]{eBOSS_DR16}
Alam S.,  et~al., 2021, \mn@doi [Phys. Rev. D] {10.1103/PhysRevD.103.083533},
  103, 083533

\bibitem[\protect\citeauthoryear{Aubourg et~al.,}{Aubourg
  et~al.}{2015}]{PhysRevD.92.123516}
Aubourg E.,  et~al., 2015, \mn@doi [\prd] {10.1103/PhysRevD.92.123516}, 92,
  123516

\bibitem[\protect\citeauthoryear{Aylor, Joy, Knox, Millea, Raghunathan  \&
  Wu}{Aylor et~al.}{2019}]{Aylor:2018drw}
Aylor K.,  Joy M.,  Knox L.,  Millea M.,  Raghunathan S.,   Wu W. L.~K.,  2019,
  \mn@doi [Astrophys. J.] {10.3847/1538-4357/ab0898}, 874, 4

\bibitem[\protect\citeauthoryear{Bahr-Kalus, Bertacca, Verde  \&
  Heavens}{Bahr-Kalus et~al.}{2021}]{Bahr-Kalus:2021jvu}
Bahr-Kalus B.,  Bertacca D.,  Verde L.,   Heavens A.,  2021, \mn@doi [JCAP]
  {10.1088/1475-7516/2021/11/027}, 11, 027

\bibitem[\protect\citeauthoryear{Bautista et~al.}{Bautista
  et~al.}{2020}]{Bautista:2020ahg}
Bautista J.~E.,  et~al., 2020, \mn@doi [Mon. Not. Roy. Astron. Soc.]
  {10.1093/mnras/staa2800}, 500, 736

\bibitem[\protect\citeauthoryear{Bernal, Verde  \& Riess}{Bernal
  et~al.}{2016}]{Bernal:2016gxb}
Bernal J.~L.,  Verde L.,   Riess A.~G.,  2016, \mn@doi [JCAP]
  {10.1088/1475-7516/2016/10/019}, 10, 019

\bibitem[\protect\citeauthoryear{Beutler et~al.,}{Beutler
  et~al.}{2011}]{Beutler:2011hx}
Beutler F.,  et~al., 2011, \mn@doi [Mon. Not. Roy. Astron. Soc.]
  {10.1111/j.1365-2966.2011.19250.x}, 416, 3017

\bibitem[\protect\citeauthoryear{Beutler et~al.,}{Beutler
  et~al.}{2012}]{Beutler:2012px}
Beutler F.,  et~al., 2012, \mn@doi [Mon. Not. Roy. Astron. Soc.]
  {10.1111/j.1365-2966.2012.21136.x}, 423, 3430

\bibitem[\protect\citeauthoryear{Blake \& Bridle}{Blake \&
  Bridle}{2005}]{Blake:2004tr}
Blake C.,  Bridle S.,  2005, \mn@doi [Mon. Not. Roy. Astron. Soc.]
  {10.1111/j.1365-2966.2005.09526.x}, 363, 1329

\bibitem[\protect\citeauthoryear{Blake et~al.}{Blake
  et~al.}{2011}]{Blake:2011rj}
Blake C.,  et~al., 2011, \mn@doi [Mon. Not. Roy. Astron. Soc.]
  {10.1111/j.1365-2966.2011.18903.x}, 415, 2876

\bibitem[\protect\citeauthoryear{Blas, Lesgourgues  \& Tram}{Blas
  et~al.}{2011}]{Blas:2011rf}
Blas D.,  Lesgourgues J.,   Tram T.,  2011, \mn@doi [JCAP]
  {10.1088/1475-7516/2011/07/034}, 07, 034

\bibitem[\protect\citeauthoryear{Box \& Cox}{Box \& Cox}{1964}]{Box:1964}
Box G. E.~P.,  Cox D.~R.,  1964, \mn@doi [Journal of the Royal Statistical
  Society: Series B (Methodological)]
  {https://doi.org/10.1111/j.2517-6161.1964.tb00553.x}, 26, 211

\bibitem[\protect\citeauthoryear{{Brieden}, {Gil-Mar{\'\i}n}  \&
  {Verde}}{{Brieden} et~al.}{2021}]{Brieden:2021edu}
{Brieden} S.,  {Gil-Mar{\'\i}n} H.,   {Verde} L.,  2021, \mn@doi [\jcap]
  {10.1088/1475-7516/2021/12/054}, \href
  {https://ui.adsabs.harvard.edu/abs/2021JCAP...12..054B} {2021, 054}

\bibitem[\protect\citeauthoryear{{Brieden}, {Gil-Mar{\'\i}n}  \&
  {Verde}}{{Brieden} et~al.}{2022a}]{BriedenTale}
{Brieden} S.,  {Gil-Mar{\'\i}n} H.,   {Verde} L.,  2022a, \mn@doi [arXiv
  e-prints] {10.48550/arXiv.2212.04522}, \href
  {https://ui.adsabs.harvard.edu/abs/2022arXiv221204522B} {p. arXiv:2212.04522}

\bibitem[\protect\citeauthoryear{{Brieden}, {Gil-Mar{\'\i}n}  \&
  {Verde}}{{Brieden} et~al.}{2022b}]{Brieden:2022lsd}
{Brieden} S.,  {Gil-Mar{\'\i}n} H.,   {Verde} L.,  2022b, \mn@doi [\jcap]
  {10.1088/1475-7516/2022/08/024}, \href
  {https://ui.adsabs.harvard.edu/abs/2022JCAP...08..024B} {2022, 024}

\bibitem[\protect\citeauthoryear{Burić \& Elezović}{Burić \&
  Elezović}{2012}]{Buric:2012}
Burić T.,  Elezović N.,  2012, \mn@doi [Integral Transforms and Special
  Functions] {10.1080/10652469.2011.591393}, 23, 355

\bibitem[\protect\citeauthoryear{{Chuang}, {Kitaura}, {Prada}, {Zhao}  \&
  {Yepes}}{{Chuang} et~al.}{2015}]{ChuangEZmock}
{Chuang} C.-H.,  {Kitaura} F.-S.,  {Prada} F.,  {Zhao} C.,   {Yepes} G.,  2015,
  \mn@doi [\mnras] {10.1093/mnras/stu2301}, \href
  {https://ui.adsabs.harvard.edu/abs/2015MNRAS.446.2621C} {446, 2621}

\bibitem[\protect\citeauthoryear{Cole et~al.}{Cole
  et~al.}{2005}]{2dFGRS:2005yhx}
Cole S.,  et~al., 2005, \mn@doi [Mon. Not. Roy. Astron. Soc.]
  {10.1111/j.1365-2966.2005.09318.x}, 362, 505

\bibitem[\protect\citeauthoryear{{Cuesta}, {Verde}, {Riess}  \&
  {Jimenez}}{{Cuesta} et~al.}{2015}]{2015MNRAS.448.3463C}
{Cuesta} A.~J.,  {Verde} L.,  {Riess} A.,   {Jimenez} R.,  2015, \mn@doi
  [\mnras] {10.1093/mnras/stv261}, \href
  {https://ui.adsabs.harvard.edu/abs/2015MNRAS.448.3463C} {448, 3463}

\bibitem[\protect\citeauthoryear{Cunnington}{Cunnington}{2022}]{Cunnington:2022ryj}
Cunnington S.,  2022, \mn@doi [Mon. Not. Roy. Astron. Soc.]
  {10.1093/mnras/stac576}, 512, 2408

\bibitem[\protect\citeauthoryear{De~Mattia et~al.}{De~Mattia
  et~al.}{2021}]{deMattia:2020fkb}
De~Mattia A.,  et~al., 2021, \mn@doi [Mon. Not. Roy. Astron. Soc.]
  {10.1093/mnras/staa3891}, 501, 5616

\bibitem[\protect\citeauthoryear{Du~Mas~des Bourboux et~al.}{Du~Mas~des
  Bourboux et~al.}{2017}]{duMasdesBourboux:2017mrl}
Du~Mas~des Bourboux H.,  et~al., 2017, \mn@doi [Astron. Astrophys.]
  {10.1051/0004-6361/201731731}, 608, A130

\bibitem[\protect\citeauthoryear{Eisenstein et~al.}{Eisenstein
  et~al.}{2005}]{SDSS:2005xqv}
Eisenstein D.~J.,  et~al., 2005, \mn@doi [Astrophys. J.] {10.1086/466512}, 633,
  560

\bibitem[\protect\citeauthoryear{Elsner, Leistedt  \& Peiris}{Elsner
  et~al.}{2016}]{Elsner:2015aga}
Elsner F.,  Leistedt B.,   Peiris H.~V.,  2016, \mn@doi [Mon. Not. Roy. Astron.
  Soc.] {10.1093/mnras/stv2777}, 456, 2095

\bibitem[\protect\citeauthoryear{{Feldman}, {Kaiser}  \& {Peacock}}{{Feldman}
  et~al.}{1994}]{FKP}
{Feldman} H.~A.,  {Kaiser} N.,   {Peacock} J.~A.,  1994, \mn@doi [\apj]
  {10.1086/174036}, \href
  {https://ui.adsabs.harvard.edu/abs/1994ApJ...426...23F} {426, 23}

\bibitem[\protect\citeauthoryear{{Ferraro} \& {Wilson}}{{Ferraro} \&
  {Wilson}}{2019}]{Ferraro:2019uce}
{Ferraro} S.,  {Wilson} M.~J.,  2019, \mn@doi [\baas]
  {10.48550/arXiv.1903.09208}, \href
  {https://ui.adsabs.harvard.edu/abs/2019BAAS...51c..72F} {51, 72}

\bibitem[\protect\citeauthoryear{{Fixsen}}{{Fixsen}}{2009}]{Fixsen}
{Fixsen} D.~J.,  2009, \mn@doi [\apj] {10.1088/0004-637X/707/2/916}, \href
  {https://ui.adsabs.harvard.edu/abs/2009ApJ...707..916F} {707, 916}

\bibitem[\protect\citeauthoryear{Gil-Marin et~al.}{Gil-Marin
  et~al.}{2020}]{Gil-Marin:2020bct}
Gil-Marin H.,  et~al., 2020, \mn@doi [Mon. Not. Roy. Astron. Soc.]
  {10.1093/mnras/staa2455}, 498, 2492

\bibitem[\protect\citeauthoryear{{Glanville}, {Howlett}  \&
  {Davis}}{{Glanville} et~al.}{2022}]{Glanville:2022xes}
{Glanville} A.,  {Howlett} C.,   {Davis} T.,  2022, \mn@doi [\mnras]
  {10.1093/mnras/stac2891}, \href
  {https://ui.adsabs.harvard.edu/abs/2022MNRAS.517.3087G} {517, 3087}

\bibitem[\protect\citeauthoryear{{Harris} et~al.,}{{Harris}
  et~al.}{2020}]{Numpy}
{Harris} C.~R.,  et~al., 2020, \mn@doi [\nat] {10.1038/s41586-020-2649-2},
  \href {https://ui.adsabs.harvard.edu/abs/2020Natur.585..357H} {585, 357}

\bibitem[\protect\citeauthoryear{{Hinton}, {Adams}, {~}  \& {Badger}}{{Hinton}
  et~al.}{2020}]{samuel_hinton_2020_4280904}
{Hinton} S.,  {Adams} C.,  {~}  {Badger} C.,  2020, {Samreay/ChainConsumer
  v0.33.0}, Zenodo, \mn@doi{10.5281/zenodo.4280904}

\bibitem[\protect\citeauthoryear{Ho, Hirata, Padmanabhan, Seljak  \&
  Bahcall}{Ho et~al.}{2008}]{Ho:2008bz}
Ho S.,  Hirata C.,  Padmanabhan N.,  Seljak U.,   Bahcall N.,  2008, \mn@doi
  [Phys. Rev.] {10.1103/PhysRevD.78.043519}, D78, 043519

\bibitem[\protect\citeauthoryear{Hojjati, Linder  \& Samsing}{Hojjati
  et~al.}{2013}]{PhysRevLett.111.041301}
Hojjati A.,  Linder E.~V.,   Samsing J.,  2013, \mn@doi [\prl]
  {10.1103/PhysRevLett.111.041301}, 111, 041301

\bibitem[\protect\citeauthoryear{Hou et~al.}{Hou et~al.}{2020}]{Hou:2020rse}
Hou J.,  et~al., 2020, \mn@doi [Mon. Not. Roy. Astron. Soc.]
  {10.1093/mnras/staa3234}, 500, 1201

\bibitem[\protect\citeauthoryear{Howlett, Lewis, Hall  \& Challinor}{Howlett
  et~al.}{2012}]{Howlett:2012mh}
Howlett C.,  Lewis A.,  Hall A.,   Challinor A.,  2012, \mn@doi [\jcap]
  {10.1088/1475-7516/2012/04/027}, 1204, 027

\bibitem[\protect\citeauthoryear{Hunter}{Hunter}{2007}]{Hunter:2007}
Hunter J.~D.,  2007, \mn@doi [Computing in Science \& Engineering]
  {10.1109/MCSE.2007.55}, 9, 90

\bibitem[\protect\citeauthoryear{Kaiser}{Kaiser}{1987}]{Kaiser:1987}
Kaiser N.,  1987, \mn@doi [Mon. Not. Roy. Astron. Soc.]
  {10.1093/mnras/227.1.1}, 227, 1

\bibitem[\protect\citeauthoryear{Kalus, Percival  \& Samushia}{Kalus
  et~al.}{2016a}]{Kalus:2015lna}
Kalus B.,  Percival W.~J.,   Samushia L.,  2016a, \mn@doi [Mon. Not. Roy.
  Astron. Soc.] {10.1093/mnras/stv2307}, 455, 2573

\bibitem[\protect\citeauthoryear{Kalus, Percival, Bacon  \& Samushia}{Kalus
  et~al.}{2016b}]{Kalus:2016cno}
Kalus B.,  Percival W.~J.,  Bacon D.,   Samushia L.,  2016b, \mn@doi [Mon. Not.
  Roy. Astron. Soc.] {10.1093/mnras/stw2008}, 463, 467

\bibitem[\protect\citeauthoryear{Kalus, Percival, Bacon, Mueller, Samushia,
  Verde, Ross  \& Bernal}{Kalus et~al.}{2019}]{Kalus:2018qsy}
Kalus B.,  Percival W.~J.,  Bacon D.~J.,  Mueller E.~M.,  Samushia L.,  Verde
  L.,  Ross A.~J.,   Bernal J.~L.,  2019, \mn@doi [Mon. Not. Roy. Astron. Soc.]
  {10.1093/mnras/sty2655}, 482, 453

\bibitem[\protect\citeauthoryear{Karamanis \& Beutler}{Karamanis \&
  Beutler}{2021}]{Karamanis:2020zss}
Karamanis M.,  Beutler F.,  2021, \mn@doi [Stat. Comput.]
  {10.1007/s11222-021-10038-2}, 31, 61

\bibitem[\protect\citeauthoryear{Karamanis, Beutler  \& Peacock}{Karamanis
  et~al.}{2021}]{Karamanis:2021tsx}
Karamanis M.,  Beutler F.,   Peacock J.~A.,  2021, \mn@doi [Mon. Not. Roy.
  Astron. Soc.] {10.1093/mnras/stab2867}, 508, 3589

\bibitem[\protect\citeauthoryear{Kazin et~al.}{Kazin
  et~al.}{2014}]{Kazin:2014qga}
Kazin E.~A.,  et~al., 2014, \mn@doi [Mon. Not. Roy. Astron. Soc.]
  {10.1093/mnras/stu778}, 441, 3524

\bibitem[\protect\citeauthoryear{Knox \& Millea}{Knox \&
  Millea}{2020}]{Knox:2019rjx}
Knox L.,  Millea M.,  2020, \mn@doi [Phys. Rev. D]
  {10.1103/PhysRevD.101.043533}, 101, 043533

\bibitem[\protect\citeauthoryear{Kobayashi, Nishimichi, Takada  \&
  Miyatake}{Kobayashi et~al.}{2022}]{Kobayashi:2021oud}
Kobayashi Y.,  Nishimichi T.,  Takada M.,   Miyatake H.,  2022, \mn@doi [Phys.
  Rev. D] {10.1103/PhysRevD.105.083517}, 105, 083517

\bibitem[\protect\citeauthoryear{{Komatsu} et~al.,}{{Komatsu}
  et~al.}{2009}]{Komatsu}
{Komatsu} E.,  et~al., 2009, \mn@doi [\apjs] {10.1088/0067-0049/180/2/330},
  \href {https://ui.adsabs.harvard.edu/abs/2009ApJS..180..330K} {180, 330}

\bibitem[\protect\citeauthoryear{Leistedt \& Peiris}{Leistedt \&
  Peiris}{2014}]{Leistedt:2014wia}
Leistedt B.,  Peiris H.~V.,  2014, \mn@doi [Mon. Not. Roy. Astron. Soc.]
  {10.1093/mnras/stu1439}, 444, 2

\bibitem[\protect\citeauthoryear{Leistedt, Peiris, Mortlock, Benoit-L\'evy  \&
  Pontzen}{Leistedt et~al.}{2013}]{Leistedt:2013gfa}
Leistedt B.,  Peiris H.~V.,  Mortlock D.~J.,  Benoit-L\'evy A.,   Pontzen A.,
  2013, \mn@doi [Mon. Not. Roy. Astron. Soc.] {10.1093/mnras/stt1359}, 435,
  1857

\bibitem[\protect\citeauthoryear{Lewis \& Bridle}{Lewis \&
  Bridle}{2002}]{Lewis:2002ah}
Lewis A.,  Bridle S.,  2002, \mn@doi [\prd] {10.1103/PhysRevD.66.103511}, 66,
  103511

\bibitem[\protect\citeauthoryear{Lewis, Challinor  \& Lasenby}{Lewis
  et~al.}{2000}]{Lewis:1999bs}
Lewis A.,  Challinor A.,   Lasenby A.,  2000, \mn@doi [\apj] {10.1086/309179},
  538, 473

\bibitem[\protect\citeauthoryear{{Macaulay} et~al.,}{{Macaulay}
  et~al.}{2019}]{2019MNRAS.486.2184M}
{Macaulay} E.,  et~al., 2019, \mn@doi [\mnras] {10.1093/mnras/stz978}, \href
  {https://ui.adsabs.harvard.edu/abs/2019MNRAS.486.2184M} {486, 2184}

\bibitem[\protect\citeauthoryear{{Mangano}, {Miele}, {Pastor}, {Pinto},
  {Pisanti}  \& {Serpico}}{{Mangano} et~al.}{2005}]{Mangano}
{Mangano} G.,  {Miele} G.,  {Pastor} S.,  {Pinto} T.,  {Pisanti} O.,
  {Serpico} P.~D.,  2005, \mn@doi [Nuclear Physics B]
  {10.1016/j.nuclphysb.2005.09.041}, \href
  {https://ui.adsabs.harvard.edu/abs/2005NuPhB.729..221M} {729, 221}

\bibitem[\protect\citeauthoryear{{Mueller} et~al.,}{{Mueller}
  et~al.}{2021}]{Mueller1}
{Mueller} E.-M.,  et~al., 2021, \mn@doi [arXiv e-prints]
  {10.48550/arXiv.2106.13725}, \href
  {https://ui.adsabs.harvard.edu/abs/2021arXiv210613725M} {p. arXiv:2106.13725}

\bibitem[\protect\citeauthoryear{{Mueller} et~al.,}{{Mueller}
  et~al.}{2022}]{Mueller2}
{Mueller} E.-M.,  et~al., 2022, \mn@doi [\mnras] {10.1093/mnras/stac812}, \href
  {https://ui.adsabs.harvard.edu/abs/2022MNRAS.514.3396M} {514, 3396}

\bibitem[\protect\citeauthoryear{Neveux et~al.}{Neveux
  et~al.}{2020}]{Neveux:2020voa}
Neveux R.,  et~al., 2020, \mn@doi [Mon. Not. Roy. Astron. Soc.]
  {10.1093/mnras/staa2780}, 499, 210

\bibitem[\protect\citeauthoryear{{Nunes}, {Vagnozzi}, {Kumar}, {Di Valentino}
  \& {Mena}}{{Nunes} et~al.}{2022}]{Nunes}
{Nunes} R.~C.,  {Vagnozzi} S.,  {Kumar} S.,  {Di Valentino} E.,   {Mena} O.,
  2022, \mn@doi [\prd] {10.1103/PhysRevD.105.123506}, \href
  {https://ui.adsabs.harvard.edu/abs/2022PhRvD.105l3506N} {105, 123506}

\bibitem[\protect\citeauthoryear{Okumura et~al.}{Okumura
  et~al.}{2016}]{Okumura:2015lvp}
Okumura T.,  et~al., 2016, \mn@doi [Publ. Astron. Soc. Jap.]
  {10.1093/pasj/psw029}, 68, 38

\bibitem[\protect\citeauthoryear{Percival et~al.}{Percival
  et~al.}{2004}]{2dFGRS:2004cmo}
Percival W.~J.,  et~al., 2004, \mn@doi [Mon. Not. Roy. Astron. Soc.]
  {10.1111/j.1365-2966.2004.08146.x}, 353, 1201

\bibitem[\protect\citeauthoryear{Peter \& Uzan}{Peter \&
  Uzan}{2013}]{Peter:2013avv}
Peter P.,  Uzan J.-P.,  2013, {Primordial Cosmology}.
Oxford Graduate Texts, Oxford University Press

\bibitem[\protect\citeauthoryear{Pezzotta et~al.}{Pezzotta
  et~al.}{2017}]{Pezzotta:2016gbo}
Pezzotta A.,  et~al., 2017, \mn@doi [Astron. Astrophys.]
  {10.1051/0004-6361/201630295}, 604, A33

\bibitem[\protect\citeauthoryear{{Philcox}, {Sherwin}, {Farren}  \&
  {Baxter}}{{Philcox} et~al.}{2021}]{Philcox}
{Philcox} O. H.~E.,  {Sherwin} B.~D.,  {Farren} G.~S.,   {Baxter} E.~J.,  2021,
  \mn@doi [\prd] {10.1103/PhysRevD.103.023538}, \href
  {https://ui.adsabs.harvard.edu/abs/2021PhRvD.103b3538P} {103, 023538}

\bibitem[\protect\citeauthoryear{Poole et~al.}{Poole
  et~al.}{2013}]{Poole:2012ex}
Poole G.~B.,  et~al., 2013, \mn@doi [Mon. Not. Roy. Astron. Soc.]
  {10.1093/mnras/sts431}, 429, 1902

\bibitem[\protect\citeauthoryear{Poulin, Smith, Karwal  \& Kamionkowski}{Poulin
  et~al.}{2019}]{Poulin:2018cxd}
Poulin V.,  Smith T.~L.,  Karwal T.,   Kamionkowski M.,  2019, \mn@doi [Phys.
  Rev. Lett.] {10.1103/PhysRevLett.122.221301}, 122, 221301

\bibitem[\protect\citeauthoryear{{Prada}, {Klypin}, {Yepes}, {Nuza}  \&
  {Gottloeber}}{{Prada} et~al.}{2011}]{Prada:2011uz}
{Prada} F.,  {Klypin} A.,  {Yepes} G.,  {Nuza} S.~E.,   {Gottloeber} S.,  2011,
  \mn@doi [arXiv e-prints] {10.48550/arXiv.1111.2889}, \href
  {https://ui.adsabs.harvard.edu/abs/2011arXiv1111.2889P} {p. arXiv:1111.2889}

\bibitem[\protect\citeauthoryear{Pullen \& Hirata}{Pullen \&
  Hirata}{2013}]{Pullen:2012rd}
Pullen A.~R.,  Hirata C.~M.,  2013, \mn@doi [Publ. Astron. Soc. Pac.]
  {10.1086/671189}, 125, 705

\bibitem[\protect\citeauthoryear{Raichoor et~al.}{Raichoor
  et~al.}{2020}]{Raichoor:2020vio}
Raichoor A.,  et~al., 2020, \mn@doi [Mon. Not. Roy. Astron. Soc.]
  {10.1093/mnras/staa3336}, 500, 3254

\bibitem[\protect\citeauthoryear{Rezaie et~al.}{Rezaie
  et~al.}{2021}]{Rezaie:2021voi}
Rezaie M.,  et~al., 2021, \mn@doi [Mon. Not. Roy. Astron. Soc.]
  {10.1093/mnras/stab1730}, 506, 3439

\bibitem[\protect\citeauthoryear{Riemer-Sørensen, Parkinson  \&
  Davis}{Riemer-Sørensen et~al.}{2013}]{riemer-sorensen_parkinson_davis_2013}
Riemer-Sørensen S.,  Parkinson D.,   Davis T.~M.,  2013, \mn@doi [\pasa]
  {10.1017/pas.2013.005}, 30, e029

\bibitem[\protect\citeauthoryear{Riess et~al.}{Riess
  et~al.}{2022}]{Riess:2021jrx}
Riess A.~G.,  et~al., 2022, \mn@doi [Astrophys. J. Lett.]
  {10.3847/2041-8213/ac5c5b}, 934, L7

\bibitem[\protect\citeauthoryear{Ross et~al.}{Ross et~al.}{2013}]{Ross:2012sx}
Ross A.~J.,  et~al., 2013, \mn@doi [Mon. Not. Roy. Astron. Soc.]
  {10.1093/mnras/sts094}, 428, 1116

\bibitem[\protect\citeauthoryear{Ross et~al.}{Ross et~al.}{2020}]{Ross:2020lqz}
Ross A.~J.,  et~al., 2020, \mn@doi [Mon. Not. Roy. Astron. Soc.]
  {10.1093/mnras/staa2416}, 498, 2354

\bibitem[\protect\citeauthoryear{Rybicki \& Press}{Rybicki \&
  Press}{1992}]{Rybicki:1992jz}
Rybicki G.~B.,  Press W.~H.,  1992, \mn@doi [Astrophys. J.] {10.1086/171845},
  398, 169

\bibitem[\protect\citeauthoryear{{Schlegel}, {Kollmeier}  \&
  {Ferraro}}{{Schlegel} et~al.}{2019}]{Schlegel:2019eqc}
{Schlegel} D.,  {Kollmeier} J.~A.,   {Ferraro} S.,  2019, in Bulletin of the
  American Astronomical Society. p.~229 (\mn@eprint {arXiv} {1907.11171}),
  \mn@doi{10.48550/arXiv.1907.11171}

\bibitem[\protect\citeauthoryear{{Sch{\"o}neberg}, {Verde}, {Gil-Mar{\'\i}n}
  \& {Brieden}}{{Sch{\"o}neberg} et~al.}{2022}]{Schoeneberg}
{Sch{\"o}neberg} N.,  {Verde} L.,  {Gil-Mar{\'\i}n} H.,   {Brieden} S.,  2022,
  \mn@doi [\jcap] {10.1088/1475-7516/2022/11/039}, \href
  {https://ui.adsabs.harvard.edu/abs/2022JCAP...11..039S} {2022, 039}

\bibitem[\protect\citeauthoryear{Scolnic et~al.}{Scolnic
  et~al.}{2018}]{Pantheon}
Scolnic D.~M.,  et~al., 2018, \mn@doi [Astrophys. J.]
  {10.3847/1538-4357/aab9bb}, 859, 101

\bibitem[\protect\citeauthoryear{{Simon}, {Zhang}, {Poulin}  \&
  {Smith}}{{Simon} et~al.}{2022}]{Simon:2022lde}
{Simon} T.,  {Zhang} P.,  {Poulin} V.,   {Smith} T.~L.,  2022, \mn@doi [arXiv
  e-prints] {10.48550/arXiv.2208.05929}, \href
  {https://ui.adsabs.harvard.edu/abs/2022arXiv220805929S} {p. arXiv:2208.05929}

\bibitem[\protect\citeauthoryear{Slosar, Seljak  \& Makarov}{Slosar
  et~al.}{2004}]{Slosar:2004fr}
Slosar A.,  Seljak U.,   Makarov A.,  2004, \mn@doi [Phys. Rev. D]
  {10.1103/PhysRevD.69.123003}, 69, 123003

\bibitem[\protect\citeauthoryear{Smith, Challinor  \& Rocha}{Smith
  et~al.}{2006}]{Smith:2005ue}
Smith S.,  Challinor A.,   Rocha G.,  2006, \mn@doi [Phys. Rev. D]
  {10.1103/PhysRevD.73.023517}, 73, 023517

\bibitem[\protect\citeauthoryear{{Smith}, {Poulin}  \& {Simon}}{{Smith}
  et~al.}{2022}]{Smith:2022}
{Smith} T.~L.,  {Poulin} V.,   {Simon} T.,  2022, \mn@doi [arXiv e-prints]
  {10.48550/arXiv.2208.12992}, \href
  {https://ui.adsabs.harvard.edu/abs/2022arXiv220812992S} {p. arXiv:2208.12992}

\bibitem[\protect\citeauthoryear{Tegmark}{Tegmark}{1997}]{Tegmark:1996qtQML}
Tegmark M.,  1997, \mn@doi [Phys. Rev.] {10.1103/PhysRevD.55.5895}, D55, 5895

\bibitem[\protect\citeauthoryear{{The MSE Science Team} et~al.,}{{The MSE
  Science Team} et~al.}{2019}]{MSEScienceTeam:2019bva}
{The MSE Science Team} et~al., 2019, \mn@doi [arXiv e-prints]
  {10.48550/arXiv.1904.04907}, \href
  {https://ui.adsabs.harvard.edu/abs/2019arXiv190404907T} {p. arXiv:1904.04907}

\bibitem[\protect\citeauthoryear{Vagnozzi, Di~Valentino, Gariazzo, Melchiorri,
  Mena  \& Silk}{Vagnozzi et~al.}{2021}]{Vagnozzi:2020rcz}
Vagnozzi S.,  Di~Valentino E.,  Gariazzo S.,  Melchiorri A.,  Mena O.,   Silk
  J.,  2021, \mn@doi [Phys. Dark Univ.] {10.1016/j.dark.2021.100851}, 33,
  100851

\bibitem[\protect\citeauthoryear{{Virtanen} et~al.,}{{Virtanen}
  et~al.}{2020}]{scipy}
{Virtanen} P.,  et~al., 2020, \mn@doi [Nature Methods]
  {10.1038/s41592-019-0686-2}, \href
  {https://ui.adsabs.harvard.edu/abs/2020NatMe..17..261V} {17, 261}

\bibitem[\protect\citeauthoryear{Wang, Percival, Avila, Crittenden  \&
  Bianchi}{Wang et~al.}{2019}]{Wang:2018xuy}
Wang M.~S.,  Percival W.~J.,  Avila S.,  Crittenden R.,   Bianchi D.,  2019,
  \mn@doi [Mon. Not. Roy. Astron. Soc.] {10.1093/mnras/stz829}, 486, 951

\bibitem[\protect\citeauthoryear{Wright et~al.}{Wright
  et~al.}{2010}]{Wright:2010qw}
Wright E.~L.,  et~al., 2010, \mn@doi [Astron. J.]
  {10.1088/0004-6256/140/6/1868}, 140, 1868

\bibitem[\protect\citeauthoryear{Y\`eche et~al.}{Y\`eche
  et~al.}{2020}]{Yeche:2020tjm}
Y\`eche C.,  et~al., 2020, \mn@doi [Res. Notes AAS] {10.3847/2515-5172/abc01a},
  4, 179

\bibitem[\protect\citeauthoryear{York et~al.}{York et~al.}{2000}]{SDSS:2000hjo}
York D.~G.,  et~al., 2000, \mn@doi [Astron. J.] {10.1086/301513}, 120, 1579

\bibitem[\protect\citeauthoryear{{Zel'dovich}}{{Zel'dovich}}{1970}]{ZeldovichApprox}
{Zel'dovich} Y.~B.,  1970, \aap, \href
  {https://ui.adsabs.harvard.edu/abs/1970A&A.....5...84Z} {5, 84}

\bibitem[\protect\citeauthoryear{{Zhai} \& {Percival}}{{Zhai} \&
  {Percival}}{2022}]{Zhai:2022zif}
{Zhai} Z.,  {Percival} W.~J.,  2022, \mn@doi [\prd]
  {10.1103/PhysRevD.106.103527}, \href
  {https://ui.adsabs.harvard.edu/abs/2022PhRvD.106j3527Z} {106, 103527}

\bibitem[\protect\citeauthoryear{{Zhao} et~al.,}{{Zhao}
  et~al.}{2021}]{ZhaoMocks}
{Zhao} C.,  et~al., 2021, \mn@doi [\mnras] {10.1093/mnras/stab510}, \href
  {https://ui.adsabs.harvard.edu/abs/2021MNRAS.503.1149Z} {503, 1149}

\makeatother
\end{thebibliography}

\appendix

\section{Alternative parameterisations}
\label{app:altparam}

The power spectrum asymptotes to a power law at small and at large scales. Thus, defining a weighting function $0\leq w(x)\leq 1\; \forall x$ such that $\lim_{x\rightarrow -\infty}m(x) = 0$ and $\lim_{x\rightarrow \infty}m(x) = 1$, we can parameterise the power spectrum at all scales as
\begin{align}
    \ln P(k) = \left[1 - m(\ln (k/k_\mathrm{TO}))\right]&\left[\ln P_a + a \ln (k/k_\mathrm{TO})\right]\nonumber\\  + m(\ln (k/k_\mathrm{TO}))&\left[\ln P_b - b \ln (k/k_\mathrm{TO})\right].
\end{align}
A candidate for $m(x)$ is $\frac{1}{2} + \frac{\arctan(cx)}{\pi}$ with an arbitrary scaling parameter $c$. This fulfils naturally $\ln P\left(k_\mathrm{TO}\right) = \frac{\ln P_a + \ln P_b}{2}$. We fix $c$ by imposing that $k_\mathrm{TO}$ is indeed the turnover scale, i.e. $P^\prime (k_\mathrm{TO}) = 0$. The non-trivial solution of this condition is 
\begin{equation}
    c = \frac{a - b}{2 \left(\ln P_a -\ln P_b\right)}.
\end{equation}
From this solution, it is evident that $P_a \neq P_b$, and, in turn, we have to marginalise over one more amplitude parameter as in the parameterisation presented in \autoref{eq:Poole_Pk}. Having done so, we do not get any stable constraints on $k_\mathrm{TO}$ and do not consider this parameterisation any further.

\section{Curvature}
\label{app:curv}

Throughout this article, we have made the ubiquitous assumption of universal flatness. The latest Planck temperature and polarisation measurements \citep{Planck:2018vyg} favour a closed universe (i.e. $\Omega_\mathrm{k} < 0$) at $2 \sigma$ when analysed without any additional external data. However, combining the same data with either BAO or Supernova data suggests a flat universe with a $1 \sigma$ accuracy of 0.2 per cent \citep{Planck:2018vyg}. Full-shape analyses of galaxy power spectra have sought to confirm or falsify the Planck results. However, different studies disagree with each other. On the one hand, \citet{Vagnozzi:2020rcz}'s full shape analysis of the BOSS DR12 CMASS power spectrum combined with Planck is consistent with spatial flatness. 
Also, \citet{Brieden:2022lsd}'s model-independent \textsc{ShapeFit} constraints from the complete set of BOSS and eBOSS data are consistent with $\Omega_\mathrm{k} = 0$.
On the other hand, using EFTofLSS to simultaneously constrain measurements from the 6dFGS, BOSS, and eBOSS catalogues, \citet{Glanville:2022xes} show a $2\sigma$ preference for curvature. 

Curvature does not affect the turnover scale. Thus our measurement of $\alpha_\mathrm{eq}$ is, in turn, independent of curvature. Any reasonable amount of curvature also becomes prominent only in late intermediate epochs of the Universe, leaving $r_\mathrm{H}$ effectively unaffected by curvature as well. What does change is
the interpretation of angles, which affects the angular diameter distance and, hence, $D_\mathrm{V}$, making it possible to distinguish predictions of a closed universe from a flat universe. 

\begin{figure}
    \centering
    \includegraphics[width = \columnwidth]{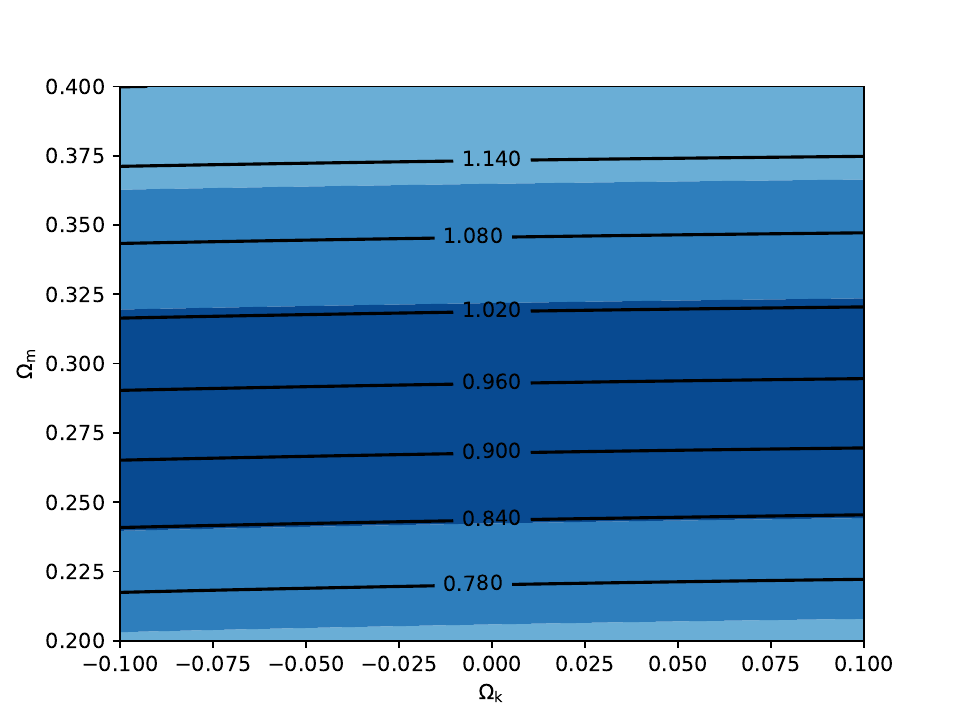}
    \caption{$\alpha_\mathrm{eq}$ as a function of $\Omega_\mathrm{m}$ and $\Omega_\mathrm{k}$ with the eBOSS $1\sigma$ and $2\sigma$ contours shaded in blue.}
    \label{fig:DV_vs_rH}
\end{figure}

In \autoref{fig:DV_vs_rH}, we show how the prediction of $\alpha_\mathrm{eq}$ changes as a function of $\Omega_\mathrm{m}h^2$ as we decrease the value of $\Omega_\mathrm{k}$. We keep $H_0$ fixed, thus, we are increasing $\Omega_\Lambda$ as we decrease $\Omega_\mathrm{k}$. In a closed universe, the sum of the angles in a triangle is less than $\pi/2$, which leads to objects observed with a certain diameter being farther, generally increasing the angular diameter distance $D_\mathrm{A}$ and, hence, $\alpha_\mathrm{eq}$. However, we observe that the prediction on $\alpha_\mathrm{eq}$ barely depends on $\Omega_\mathrm{k}$ and, in turn, the turnover measurement at low-redshift by itself (without some high-redshift anchor from the CMB) is not a good probe of curvature.

\bsp	
\label{lastpage}
\end{document}